\documentclass[journal]{IEEEtran}

\ifCLASSINFOpdf
\else
\fi

\usepackage{xcolor,colortbl}
\usepackage{adjustbox}
\usepackage{graphicx}
\usepackage{mathtools}
\usepackage{gensymb}
\usepackage{subfigure}
\usepackage{pbox}
\usepackage{multirow}
\usepackage{easyReview}
\usepackage{lipsum}  
\usepackage{url}  
\usepackage{tabularx}
\usepackage{subfigure}
\usepackage{longtable}
\usepackage{supertabular,booktabs}

\definecolor{languidlavender}{rgb}{0.84, 0.79, 0.87}
\begin{document}

\title{A Review on Deep-Learning Algorithms for Fetal Ultrasound-Image Analysis}

\author{Maria Chiara~Fiorentino, Francesca Pia~Villani, Mariachiara Di~Cosmo,
        Emanuele Frontoni,
        and~Sara~Moccia
\thanks{M. C. Fiorentino and M. Di Cosmo are with the Department of Information Engineering, Università Politecnica delle Marche (Italy) }
\thanks{F. P. Villani is with the Department of Humanities, Universit{à} degli Studi di Macerata (Italy)}
\thanks{E. Frontoni is with and the Department of Information Engineering, Università Politecnica delle Marche and the Department of Political Sciences, Communication and International Relations, Università degli Studi di Macerata (Italy)}
\thanks{S. Moccia is with the Biorobotics Institute and Department of Excellence in Robotics \& AI, Scuola Superiore Sant'Anna (Italy).}
}

\maketitle

\begin{abstract}
Deep-learning (DL) algorithms are becoming the standard for processing ultrasound (US) fetal images.
{Despite the large number of survey papers already present in this field, most of them are focusing on a broader area of medical-image analysis or not covering all fetal US DL applications.} This paper surveys the most recent work in the field, with a total of 145 research papers published after 2017. Each paper is analyzed and commented from both the methodology and application perspective. We categorized the papers in (i) fetal standard-plane detection, (ii) anatomical-structure analysis and (iii) biometry parameter estimation. For each category, main limitations and open issues are presented. Summary tables are included to facilitate the comparison among the different approaches. 
Publicly-available datasets and performance metrics commonly used to assess algorithm performance are summarized, too. 
This paper ends with a critical summary of the current state of the art on DL algorithms for fetal US image analysis and a discussion on current challenges that have to be tackled by researchers working in the field to translate the research methodology into the actual clinical practice. 
\end{abstract}


\IEEEpeerreviewmaketitle

\section{Introduction}
\label{sec:intro}

\IEEEPARstart{U}{ltrasound} (US) imaging is an imaging modality widely used for the diagnosis, screening and treatment of a large number of diseases, due to its portability, low cost and non-invasive nature \cite{zaffino2020review}.
In the years, US imaging has turned out to be the preferred checkup method during pregnancy~\cite{whitworth2015ultrasound,dias2014ultrasound}. It is commonly used to evaluate fetus's growth and development, as well as to monitor pregnancy and assess clinical suspicion~\cite{bijma2008decision}.
\begin{figure*}[tbp]
    \centering
    \includegraphics[width = .8\textwidth]{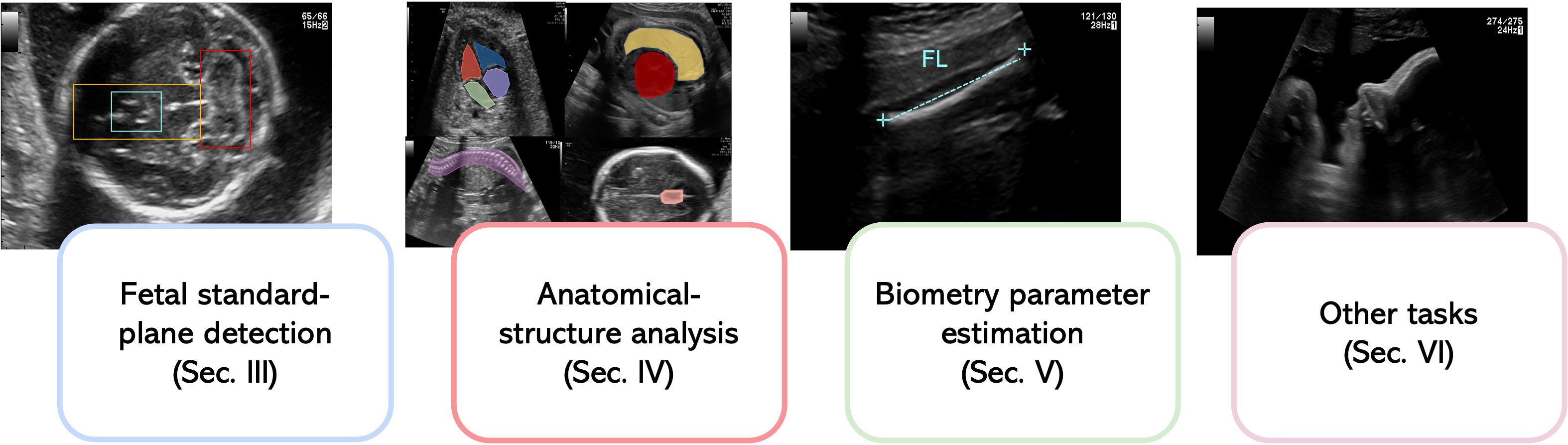}
    \caption{Summary of the tasks surveyed in this paper. }
    \label{fig:summary}
\end{figure*}

From the clinician perspective, analysing US images may be challenging due to the presence of artifacts, such as acoustic shadows, speckle noise, motion blurring and missing boundaries, which are produced as the result of the complex interaction between US waves and mother and fetal biological tissues~\cite{meng2020automatic}. 

During the last decades, deep learning (DL), and in particular convolutional neural networks (CNNs), have undergone an increasing role in fetal US image analysis to offer decision support to clinicians, and today an extensive literature exists.
%
Survey papers in the field have been published in the last years, even if a number of them focuses on the broader area of medical-image analysis \cite{liu2019deep} or surveys also DL algorithms for US image analysis outside the fetal field \cite{van2019deep,akkus2019survey,ouahabi2021deep,shen2021artificial,zaffino2020review}. 
Survey papers specifically dealing with US fetal images include: \cite{torrents2019segmentation,song2021classification,sree2019ultrasound}, where segmentation and classification algorithms are covered; \cite{garcia2020machine,morris2021deep} that survey methods for fetal cardiology images; \cite{rawat2018automated,bushra2021obstetrics} that briefly summarize DL methods for fetal abnormality detection, and~\cite{chen2021artificial} that analyzes research papers from a clinical perspective.


An updated review that surveys the most recent work in the field of fetal US image analysis with DL could be a valuable and compact source of information for young researchers, and a reference overview document for those already working in the field.
With this aim, our review starts describing publicly-available datasets in the field, as well as commonly used metrics for algorithm performance assessment (Sec. \ref{sec:dataset}). As shown in Fig.~\ref{fig:summary}, research papers ranging from fetal standard-plane detection (Sec.~\ref{sec:plane_det}) to anatomical-structure analysis (Sec.~\ref{sec:organs}) and fetal-biometry estimation (Sec.~\ref{sec:biometry}) are surveyed. {These sections mirror the steps of what is currently done in clinical practice to evaluate fetus well-being. A miscellaneous section (Sec.~\ref{sec:other}) is also included, collecting papers on emerging tasks, from less common fetus evaluation applications to probe movement control.} 
For each section, methods are described highlighting pros and cons. Limitations and open issues are further discussed. Summary tables are included to report information on training and testing datasets, as well as achieved performance.
A discussion on future directions and open challenges in the field of fetal US analysis with DL concludes this review (Sec. \ref{sec:disc}).

\subsection{Survey strategy}

Our survey strategy started with the following research questions:
\begin{itemize}
    \item Which are the most investigated tasks addressed using DL in the field of fetal US image analysis?
    \item Which are the main challenges in regard to fetal examination that are currently tackled by using DL? 
    \item Are the commonly-used datasets sufficient enough for robust DL algorithm development and testing?
    \item Which are 
    open issues that still have to be addressed by DL in the field?
\end{itemize}

With these questions in mind, we outlined a set of keywords for our survey, including: \textit{classification}, \textit{detection}, \textit{segmentation}, \textit{fetal}, \textit{ultrasound}, \textit{deep learning} combined together with terms related to fetal examination and organs.
The research databases were IEEEXplore, Scopus, SpringerLink, Sciencedirect and Pubmed. For each resulting paper, an extensive review of its reference list was performed. 
To better focus on the most recent and interesting trends and to not overlap with previous review work, we analysed indexed journal and conference papers published from 2017.

\begin{figure*}
     \centering
     \subfigure[]{\label{fig:a}\includegraphics[width=.35\textwidth]{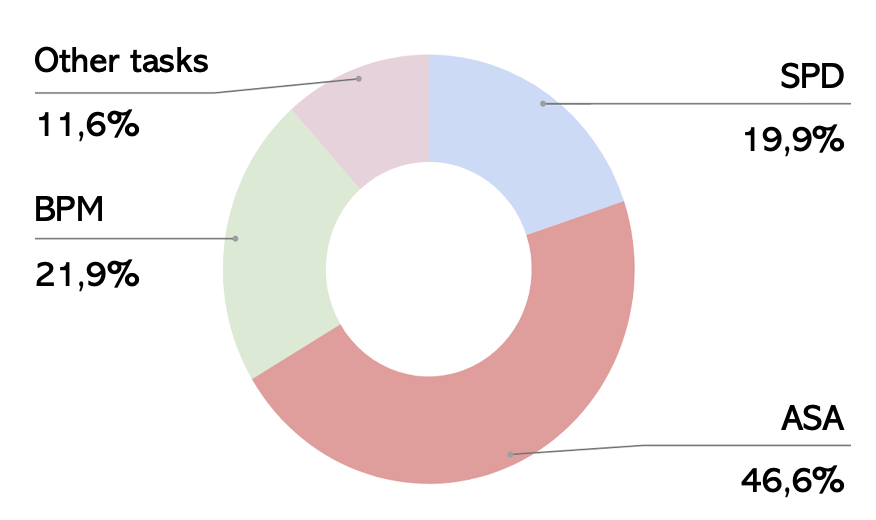}}
     \subfigure[]{\label{fig:b}\includegraphics[width=.32\textwidth]{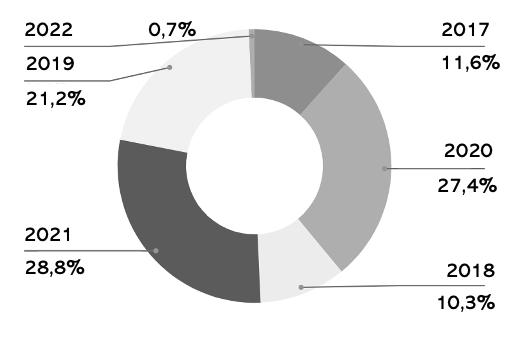}}
    \subfigure[]{\label{fig:c}\includegraphics[width=.33\textwidth]{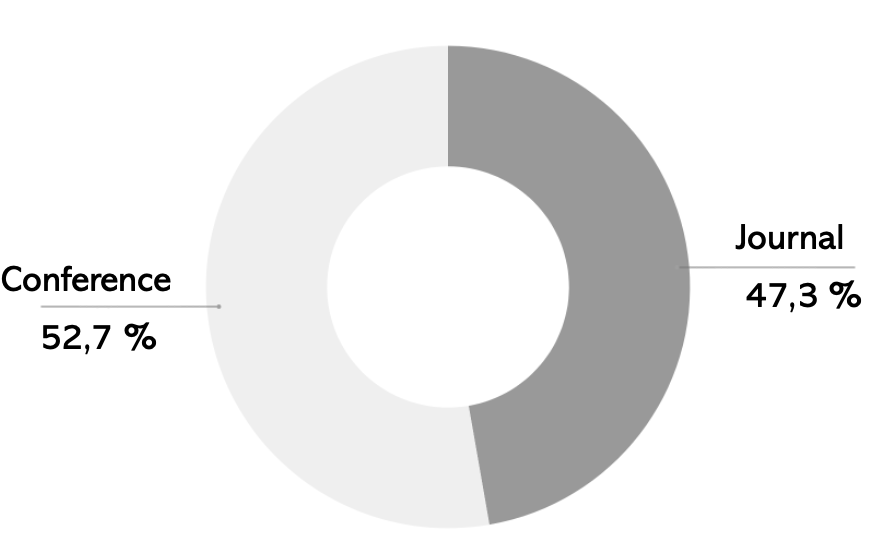}}
     \subfigure[]{\label{fig:d}\includegraphics[width=.35\textwidth]{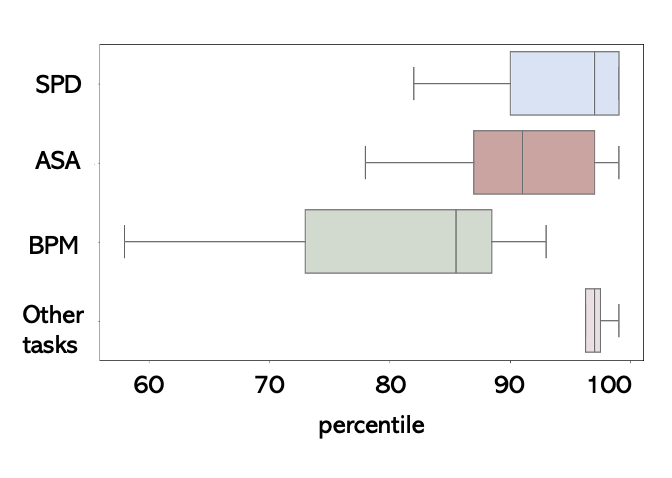}}

     \caption{\label{fig:statistics} Summary of the papers included in this paper considering: (a) addressed task, (b) year of publication, (c) contribution (journal/ conference), (d) percentile according to CiteScore rank 2020 (only for journal papers). SPD = standard-plane detection, ASA = anatomical-structure analysis, BPM = biometry parameter estimation
     }
\end{figure*}

A summary of the characteristics of the papers analyzed in this review is shown in Fig. \ref{fig:statistics}. 

\section{
Performance metrics and Publicly-available datasets}

\label{sec:dataset}

This section presents the quantitative metrics used to assess DL algorithm performance as well as publicly-available datasets for algorithm development and testing. 

\subsection{Performance metrics}

\begin{table}[tbp]
\caption{Contingency table.}
\label{Cont_table}
\centering
\begin{tabular}{cccc}
\cline{3-4}
\multicolumn{1}{ c }{} & \multicolumn{1}{ c| }{} & \multicolumn{2}{ c |}{ Gold Standard}\\
\cline{3-4}
\multicolumn{1}{ c}{} & \multicolumn{1}{ c| }{} & \multicolumn{1}{ c| }{\hphantom{on}Yes\hphantom{on}} & \multicolumn{1}{ c |}{ No}\\
\hline
\multicolumn{1}{| c}{Algorithm} & \multicolumn{1}{ |c| }{Yes} & \multicolumn{1}{ c }{$TP$} & \multicolumn{1}{ c |}{$FP$}\\
\cline{2-2}
\multicolumn{1}{ |c}{result} & \multicolumn{1}{ |c| }{No} & \multicolumn{1}{ c }{$FN$} & \multicolumn{1}{ c|}{$TN$}\\
\hline
\end{tabular}
\end{table}

\setlength{\tabcolsep}{6pt} 
\renewcommand{\arraystretch}{1.5} 

\begin{table}[tbp]
\caption{Performance metrics.}
\label{table:measures}

\begin{tabularx}{.5\textwidth}{XXX}
Index & Description  \\
\hline
 Accuracy ($Acc$)  &  ${\frac{TP + TN}{N}}$  \\
Recall ($Rec$) & ${\frac{TP}{TP + FN}}$  \\
 Specificity ($Spec$) & ${\frac{TN}{TN + FP}}$  \\
 Precision ($Prec$) & ${\frac{TP}{TP + FP}}$ \\
 F1-score ($F1$) & ${\frac{2*Prec*Spec}{Prec + Spec}}$\\
top-1 error rate / top-3 error rate  &  Top-1 and top-3 errors \\
AUC & Area Under the Receiver Operating Characteristic curve \\ 
AUC by Judd  &  AUC variant from Judd et al. \cite{judd2009learning}  \\ 
Dice similarity coefficient ($DSC$) &  Eq. \ref{dsc}   \\
Intersection Over Union ($IoU$)  &  Eq. \ref{iou}  \\ 
Hausdorff distance ($HD$)  &  Eq. \ref{hd}  \\ 
mean Average precision ($mAP$)  &  Eq. \ref{map}  \\ 
Mean absolute error ($MAE$) &  Eq. \ref{mae}  \\
Difference ($DF$) &  Eq. \ref{df} \\
Mean squared error ($MSE$) &  Eq. \ref{mse}  \\
Euclidean distance ($ED$) &  Eq. \ref{eucl}  \\
Kullback-Leiber divergence ($D_{KL}$)  &  Eq. \ref{kl}  \\ 
Normalized Scanpath Saliency ($NSS$) & Eq. \ref{nss} \\
\end{tabularx}
\end{table}

Fetal US DL algorithms are evaluated using different metrics according to the addressed task. 
For classification tasks, performance is assessed by means of the contingency table (Table \ref{Cont_table}), with True Positives ($TP$), True Negatives ($TN$), False Negatives ($FN$) and False Positives ($FP$). 
Popular classification metrics computed from the contingency table are: (1) Accuracy ($Acc$) = number of correct predictions ($TP$ + $TN$) divided by the total number of predictions ($N$); (2) Recall ($Rec$) = fraction of actual positives which are correctly identified; (3) Specificity ($Spec$) = fraction of actual negatives which are correctly identified; (4) Precision ($Prec$) = proportion of positives which are identified. The F1-score ($F1$) metric, which is equal to the harmonic mean of $Prec$ and $Rec$, is also often adopted, especially for imbalanced datasets.
These metrics can be computed at patch, image or patient level.

Other popular metrics are top-1 and top-3 error rates. For top-1 error rate, only the top class (i.e., the one with the highest output probability) is compared with the target label. In the case of top-3 error rate, the target label is compared with the top 3 predictions (i.e., the 3 with the highest probability). In both cases, the top score is computed as the number of times the predicted label matched the target label, divided by $N$. 

The receiver operating characteristic (ROC) curve, which shows the performance of a binary classifier as function of its cut-off threshold, is often reported. The area (AUC) under the ROC is used as metric and is interpreted as the probability that the DL model ranks a random positive example more highly than a random negative example. The higher the AUC (close to 1), the better the model performance.
\\

\setlength{\tabcolsep}{6pt} 
\renewcommand{\arraystretch}{1.5} 

\begin{table*}[tbp]
    \centering
        \caption{\label{table:datasets} Publicly available datasets for the development and testing of deep-learning algorithms for fetal-ultrasound image analysis. For performance metrics refer to Table \ref{table:measures}. \textbf{*}: for the HC18 challenge dataset, only the total number of subjects (551) is reported. MVP: maximum vertical pocket.}
    \begin{tabularx}{\textwidth}{XXXXlX}

    \vspace{-.1cm}
         Name   &  Task(s) & Training set size & Testing set size   & Annotator(s) & Performance metrics \\
               &   & (images/ patients) & (images/ patients)  &  \\
         \hline
         
         HC18 challenge dataset (2018) & Head-circumference estimation & 999 2D/ \textbf{*} & 335 2D/ \textbf{*}  & 2 
         & $ADF$ [mm], $DF$ [mm], $DSC$ [\%], HD [mm]\\
         
         A-AFMA ultrasound challenge dataset (2021) & 1) Amniotic fluid and maternal bladder detection
         2) MVP detection & - & - & - & $mAP$ [0-1]\\ 
         
         Burgos et al., \cite{burgos2020evaluation} (2020) & Standard plane detection of abdomen, brain, heart, femur, maternal cervix and other & 7129 2D/ 896 & 5271 2D/ 896 & 1 
         & top-1 error rate [\%], top-3 error rate [\%], $Acc$ [\%]  \\

    \end{tabularx}
\end{table*}


As regard segmentation tasks, model performance is commonly evaluated by means of the Dice similarity coefficient ($DSC$)  and Intersection over Union ($IoU$). $DSC$ is equivalent to $F1$ and can be defined as:
\begin{equation}
\label{dsc}
DSC = \frac{2|X \cap Z|}{|X| + |Z|} = \frac{2TP}{FP+FN+2TP} = F1
\end{equation}
where $X$ and $Z$ are the predicted and the labeled masks, respectively. 
The $IoU$ is defined as:
\begin{equation}
\label{iou}
IoU = \frac{|X \cap Z|}{|X \cup Z|}
\end{equation}
The two metrics are positively correlated, however $IoU$ tends to penalize single instances of bad segmentation more than $DSC$, which tends to measure something closer to average performance.

The Hausdorff distance ($HD$), which measures how far the labeled segmentation mask is from the predicted segmentation, can be used, too:
\begin{equation}
\label{hd}
HD(X,Z) = max(h(X,Z),h(X,Z))
\end{equation}
where 
\begin{equation}
h(X,Z) = \underset{x \in X}{max}\underset{z \in Z}{max} ||x-z||
\end{equation}
\\

For localization tasks, performance is assessed by means of the average precision ($AP$), which is the $Prec$ averaged across all $Rec$ values between 0 and 1. $AP$ may be seen as the area under the precision-recall curve. The $Prec$ and $Rec$ are commonly computed at various thresholds of $IoU$ between the predicted and labeled bounding box. 
For  detection tasks, where multiple structures have to be detected in the image, mean $AP$ ($mAP$) is used: 
\begin{equation}
\label{map}
mAP = \frac{\sum_{i=1}^{K}AP_i}{K}
\end{equation}
where $K$ is the number of classes.

For regression tasks, where a numerical value has to be predicted (e.g., for biometry estimation), popular metrics are:
\begin{itemize}

\item Mean squared error ($MSE$), which computes the average squared error between the predicted ($y_i$) and actual values ($\hat{y_i}$): 
\begin{equation}
\label{mse}
MSE = \frac{1}{N}\sum_{i=1}^{N}(y_i - \hat{y_i})^2
\end{equation}
Due to its differentiable nature, it is often used as loss function during training.

\item Mean absolute error ($MAE$), sometimes referred to as mean absolute deviation, defined as the average of the absolute distance between $y_i$ and $\hat{y_i}$:
\begin{equation}
\label{mae}
MAE = \frac{1}{N}\sum_{i=1}^{N}|y_i - \hat{y_i}|
\end{equation}
$MAE$ is reported along with $MSE$ because it is more robust to outliers.

\item Difference error ($DF$):
\begin{equation}
\label{df}
DF = \frac{1}{N}\sum_{i=1}^{N}(y_i - \hat{y_i})
\end{equation}

\item Euclidean distance ($ED$):
\begin{equation}
\label{eucl}
ED = \frac{1}{N}\sum_{i=1}^{N}\sqrt{(y_i - \hat{y_i})^2}
\end{equation}
\end{itemize}

Regression metrics can be also used to evaluate mask contours obtained with segmentation algorithms.
\\


Recently, researchers are focusing on visual saliency prediction (i.e., predicting human eye fixations on images in the form of a saliency maps). Models for saliency prediction can be evaluated using a variety of performance metrics, as described in \cite{bylinskii2018different}.
Among these, in the field of US fetal image analysis the Kullback-Leiber divergence ($D_{KL}$) is often used, which is defined as:
%
\begin{equation}
\label{kl}
D_{KL}(S||F^{D}) = \sum_{i=1}^{T}F^{D}_i log (\epsilon + \frac{F^{D}_i}{\epsilon + S_i})
\end{equation}
where $S$ and $F^{D}$ are the (predicted) saliency map and (ground truth) continuous fixation map distribution, respectively, $\epsilon$ is a regularization constant and $T$ is the total number of fixated pixels.

The Normalized Scanpath Saliency ($NSS$) \cite{peters2005components} is used, too. The $NSS$ is computed as:
\begin{equation}
\label{nss}
NSS(S,F^{B}) = \frac{1}{T} \sum_{i=1}^{T}\overline{S_i}  F_{i}^{B}
\end{equation}
where $F^{B}$ is the ground truth (binary) fixation location map and
\begin{equation}
\overline{S}= \frac{S - \mu(S)}{\sigma(S)} 
\end{equation}

An AUC variant from Judd et al. \cite{judd2009learning} is also used. Here the AUC is built considering $TP$ and $FP$ rates defined as follows. For a given threshold, the $TP$ rate is the ratio of true positives to the total number of fixations, where true positives are saliency map values above threshold at fixated pixels. 
The $FP$ rate is the ratio of false positives to the total number of saliency map pixels at a given threshold, where false positives are saliency map values above threshold at unfixated pixels. 


Table \ref{table:measures} summarizes the performance metrics described in this section.


\begin{figure}[tbp]
     \centering
         \includegraphics[width=.48\textwidth]{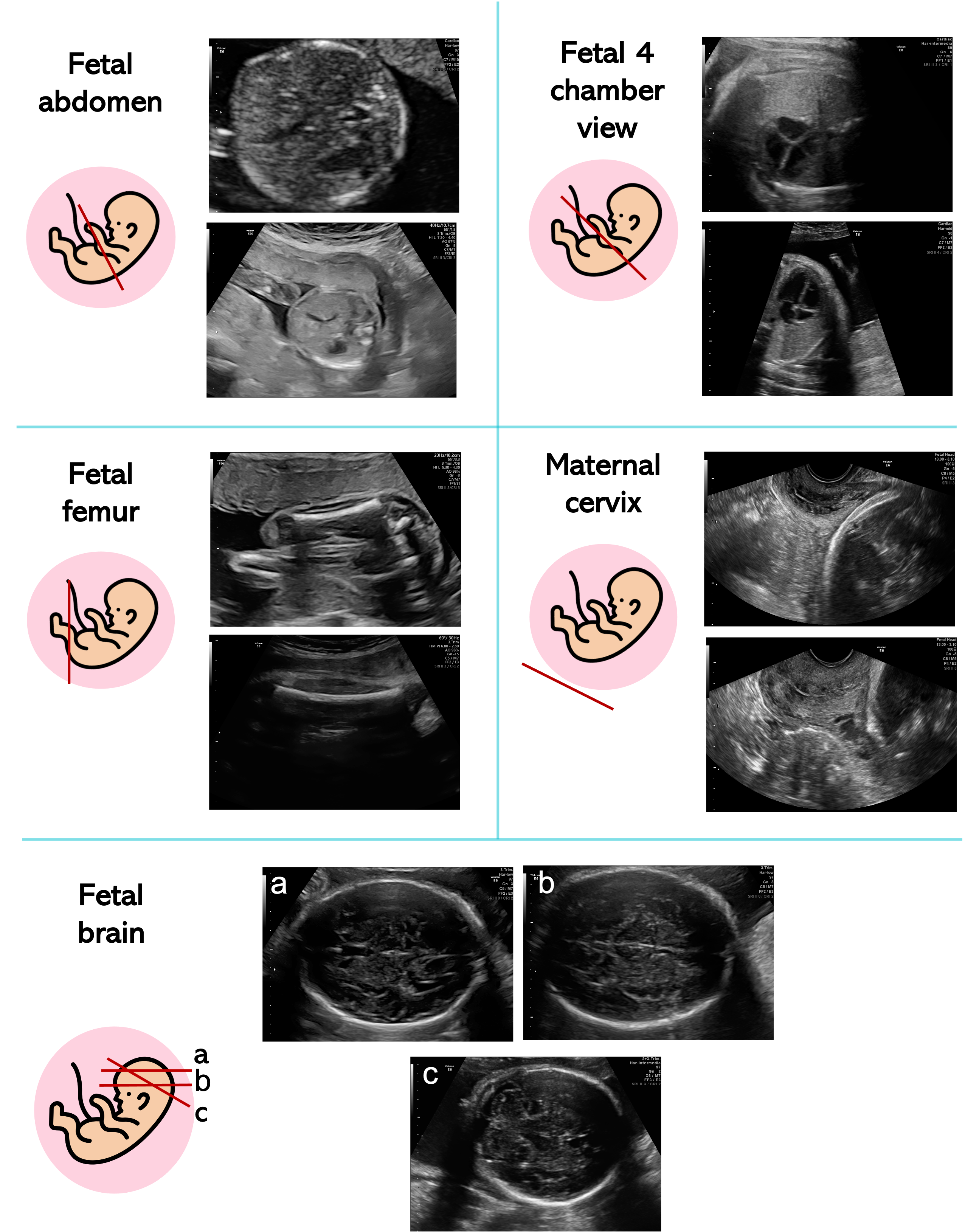} 
     \caption{\label{fig:fplanes} Visual samples of the most common fetal standard planes. }
\end{figure}



\subsection{Publicly-available datasets}

Collecting and sharing high-quality annotated fetal US datasets is not trivial. 
Labeling large datasets can take a significant amount of time, which may vary according to the task (e.g., pixel-level labeling for segmentation is way more time-consuming than image-level labeling for classification). 
Data privacy and protection concerns further constitute a barrier to data sharing. 
To attenuate these issues, international scientific organizations are working to collect and share publicly available databases to encourage algorithm development and fair comparison among algorithms. 
In the framework of the IEEE International Symposium on Biomedical Imaging (ISBI) and the  International Conference on Medical Image Computing and Computer Assisted Intervention (MICCAI), three annotated datasets have been released in the form of a Grand Challenge\footnote{https://grand-challenge.org/challenges/}.

The first challenge in time was the Challenge US: Biometric Measurements from Fetal Ultrasound Images~\cite{rueda2013evaluation}, held in conjunction and with the support of ISBI in 2012. The goal of the challenge was to automatically segment fetal abdomen, head and femur for measuring standard biometric parameters. 
Despite the recognized value of the dataset, its size (270 images) does not allow researchers to develop generalizable DL algorithms. 

In 2018, with the release of the HC18 challenge dataset\footnote{https://saras-mesad.grand-challenge.org/}~\cite{van2018automated}, the potential of DL for biomerty parameter estimation has been unlocked. The challenge was to develop algorithms to automatically measure fetal head circumference (HC), with 999 and 335 2D US images for training and testing, respectively. Images were annotated by an experienced sonographer and a medical researcher. 

During ISBI 2021, the A-AFMA ultrasound challenge\footnote{https://a-afma.grand-challenge.org/} was organized. The goals were to: 1) detect amniotic fluid and maternal bladder, 2) identify the appropriate landmarks for maximum vertical pocket (MVP) measurement, as to assess amniotic fluid volume. 

Recently, a large dataset\footnote{https://zenodo.org/record/3904280} of routinely acquired maternal-fetal screening US images was made publicly available in~\cite{burgos2020evaluation}. It consists of 7129 2D training images from 896 patients, categorized into 6 classes: abdomen, brain, femur, thorax, maternal cervix and other. Images were manually labeled by an expert fetal clinician. A test set, which consists of 5271 2D images from 896 patients, was provided. The goal was to boost the research in the field of fetal standard-plane detection.

The publicly-available datasets for the development of DL algorithms for fetal US are summarized in Table~\ref{table:datasets}





\setlength{\tabcolsep}{6pt} 
\renewcommand{\arraystretch}{1.5} 

\begin{table*}[tbp]
\scriptsize
\centering

    \caption{Summary of deep-learning algorithms for fetal standard-plane detection (for performance metrics refer to Table \ref{table:measures}). FASP = Fetal Abdomen Standard Plane, FFSP = Fetal Facial Standard Plane, FFASP = Fetal Face Axial Standard Plane, 4CH = Four Chamber View, FBSP = Fetal Brain Standard Plane, LVOT = Left Ventricular Outflow Tract, 3VV = Three-Vessel View, RVOT = Right Ventricular Outflow Tract, FFESP = Fetal Femur Standard Plane, FTSP = Fetal Trans-Thalamic Standard Plane, FCSP = Fetal Trans-Cerebellum Standard Plane, FLVSP= Fetal Lumbosacral Spine Standard Plane, FVSP = Fetal Trans-Ventricular Standard Plane. Performance metrics are expressed as the average value over the classes. *: Only the total number of subjects is reported. 
    }\label{table:plane detection}
    \begin{tabularx}{\textwidth}{XXXXlX} 

    Paper (Year) & Plane & Training set size & Test set size  & Annotators & Performance metrics \\
    \hline
    
    Wu et al.,  \cite{wu2017fuiqa} (2017) & FASP & 8072 2D (492 videos) (-) & 2606 2D (219 videos) (66 subjects) & 3 & AUC=0.99, $Acc$=0.98, $Rec$=0.96, $Spec$=0.97\\
    
    Yu et al., \cite{yu2017deep} (2017) & FFSP & 4849 2D & 2418 2D & Few & AUC=0.99, $Acc$=0.96, $Prec$=0.96, $Rec$=0.97, $F1$=0.97\\
    
    Qu et al., \cite{qu2020standard} (2020) & FBSPs & 15314 2D (155 subjects*) & 3828 2D (155 subjects*) & Few & $Acc$=0.93, $Prec$=0.93, $Rec$=0.92, $F1$=0.93\\
     
    Burgos-Artizzu et al., \cite{burgos2020evaluation} (2020) & Multiple & 7129 2D (896 subjects*) & 5271 2D (896 subjects*) & 1 &  6.2\% top-1 error, 0.27\% top-3 error, $Acc$ = 0.94 \\
   
    Kong et al., \cite{kong2018automatic} (2018) & 4CH, FASP, FBSP, FFSPs & 17036 2D & 5678 2D & 1 & $Prec$ = 0.98, $Rec$ = 0.98, $F1$ = 0.98 \\
    
    Liang et al., \cite{liang2019sprnet} (2019) & 4CH, FASP, FBSP, FFASP, coronal FFSP & 17840 2D & 4455 2D & - & $Acc$ = 0.99, $Rec$ = 0.96, $Spec$ = 0.99, $F1$ = 0.95\\ 
    
    Sundaresan et al., \cite{sundaresan2017automated} (2017) & 4CH, LVOT, 3VV & 10000 2D (10 subjects) & 2178 2D (2 subjects) & 1 & error rate = 0.23 \\

    Meng et al., \cite{meng2020unsupervised} (2020) & 4CH, FASP, LVOT, RVOT, FFESP, Lips & 12000 2D &  5500 2D & Few & $F1$=0.77, $Rec$=0.77, $Prec$= 0.78\\
    
    Montero et al., \cite{montero2021generative} (2021) & FBSP & 6498 2D & 2249 2D & Few & $Acc$ = 0.81, AUC = 0.86, $F1$ = 0.80\\ 
    
    Chen et al., \cite{chen2017ultrasound} (2017) & FASP, FFASP, 4CH & 37376 2D (900 subjects) &  13248 2D (331 subjects) & 1 & $Acc$=0.87, $Prec$=0.71, $Rec$=0.64, $F1$=0.64\\
    
    Pu et al., \cite{pu2021automatic} (2021) & FASP, FTSP, FCSP, FLVSP & 68296 2D (1199 videos) & 16740 2D (244 videos) & Few & $Acc$ = 0.85, $Prec$ = 0.85, $Rec$ = 0.85, $F1$ = 0.85\\ 
    
    Schlemper et al., \cite{schlemper2019attention} (2019) & Multiple & 122233 2D & 38243 2D  & Few & $Acc$ = 0.98, $Rec$ = 0.93, $Prec$ = 0.93, $F1$ = 0.93 \\
     
    Cai et al., \cite{cai2018multi} (2018) & FASP & 1292 2D (25 subjects) & 324 2D (8 subjects) & Few & $Prec$ = 0.96, $Rec$ = 0.96, $F1$ = 0.96 \\
    
    Cai et al., \cite{cai2020spatio} (2020) & FASP, FBSP, FFESP  &  5-fold cross validation on 280 videos & 5-fold cross validation on 280 videos & 1 & $Prec$ = 0.98, $Rec$ = 0.85, $F1$ = 0.87 \\ 
    
    Lee et al., \cite{lee2021principled} (2021) & Multiple & 1504 2D & 752 2D & - & $Prec$ = 0.75, $Rec$ = 0.73, $F1$ = 0.74\\ 
    
    Dong et al., \cite{dong2019generic} (2019) & 4CH & 5626 2D & 1406 2D & Few &  $mAP$ = 0.81\\ 
     
    Lin et al., \cite{lin2018quality} (2018) & FBSP & 4800 2D & 1153 2D & 1 & $AP$=0.79, $Rec$= 0.85,$Prec$= 0.87 \\
    
    Lin et al., \cite{lin2019multi} (2019) & FBSP & 1451 2D & 320 2D & Few & $mAP$= 0.93\\
    
    Zhang et al., \cite{zhang2021automatic} (2021) & FASP, FBSP, 4CH & 2460 2D & 820 2D & 2 & $mAP$=0.95, $Acc$=0.95, $Prec$= 0.95, $Rec$=0.93\\ 
     
    Baumgartner et al., \cite{baumgartner2017sononet} (2017) & Multiple & 140827 2D and 2438 videos (2694 subjects*) & 109165 2D and 200 videos (2694 subjects*) & 45 &  $Prec$ = 0.77, $Rec$ = 0.90 $F1$ = 0.80 and $IoU$ = 0.62 \\ 
        
    Yaqub et al., \cite{yaqub2017deep} (2017) & FVSP & 15870 2D & 3968 2D & Few & $Acc$=0.95 \\ 
    
    Gao et al., \cite{gao2020label} (2020) & FTSP, FCSP & 34586 2D from 441 videos (147 subjects) & 60 videos (20 subjects) & 2 & $mAP$=0.87 \\
    
    Chen et al., \cite{chen2019self} (2019) & Multiple & 140827 2D and 2438 videos (2694 subjects*) & 109165 2D and 200 videos (2694 subjects*) & 45 & $Prec$=0.89, $Rec$=0.90, $F1$=0.89 \\ 
    
    Tan et al., \cite{tan2019semi} (2019) & Multiple & 22757 2D (2000 subjects*) & 5737 2D (2000 subjects*) & - & $Acc$=0.70 \\
    
    Dou et al., \cite{dou2019agent} (2019) & FBSPs & 330 3D (330 subjects) & 100 3D (100 subjects) & 1 & $DF$=3.03 mm, $\theta$=9.36$\degree$  \\
        
    Yang et al., \cite{yang2021agent} (2021) &  Multiple & 1281 3D & 354 3D & 4 & $DF$=2.31 mm, $\theta$=10.36\degree \\
    
    Yang et al., \cite{yang2021searching} (2021) & Uterine standard planes & 539 3D (476 subjects*) & 144 3D (476 subjects*) & 4 & $DF$=1.82 mm, $\theta$=7.20$\degree$ \\ 
    
    Li et al., \cite{li2018standard} (2018) & FVSP, FCSP & 50 3D (50 subjects) & 22 3D (22 subjects) & 1 & mean plane centre difference = 3.44 mm, rotation angle = 11.05$\degree$ \\ 
    
    Tsai et al., \cite{tsai2021automatic} (2021) & Middle sagittal plane & 112 3D & 28 3D & - & $ED$ = 0.05 \\

    \end{tabularx}
\end{table*}

\section{Fetal standard-plane detection}
\label{sec:plane_det}

According to the International Society of Ultrasound in Obstetrics and Gynecology (ISUOG) guidelines \cite{salomon2019isuog}, the use of standardized planes of acquisition improves the reproducibility of (i) fetal biometry assessment and (ii) fetus evaluation. 
%
%
The planes that are typically acquired to extrapolate biometric measurements are  fetal abdomen (FASP), brain (FBSP) and femur (FFESP) standard planes. FBSP involves trans-ventricular (FVSP) and trans-thalamic (FTSP) standard planes.
Fetus evaluation further requires the acquisition of maternal cervix, fetal heart (including 4 chamber view (4CH), left ventricular outflow tract (LVOT), right ventricular outflow tract (RVOT), three-vessel trachea (3VT), three-vessel view (3VV)), fetal trans-cerebellum standard plane (FCSP), fetal facial standard plane (FFSP) and lumbosacral spine plane (FLVSP). FFSP includes axial (FFASP), coronal and sagittal planes. 
Visual samples of the most common standard planes are shown in Fig.~\ref{fig:fplanes}.

\begin{figure}[tbp]
\centering
\subfigure[\label{subfig:class} Classification]
   {\includegraphics[width=.26\textwidth]{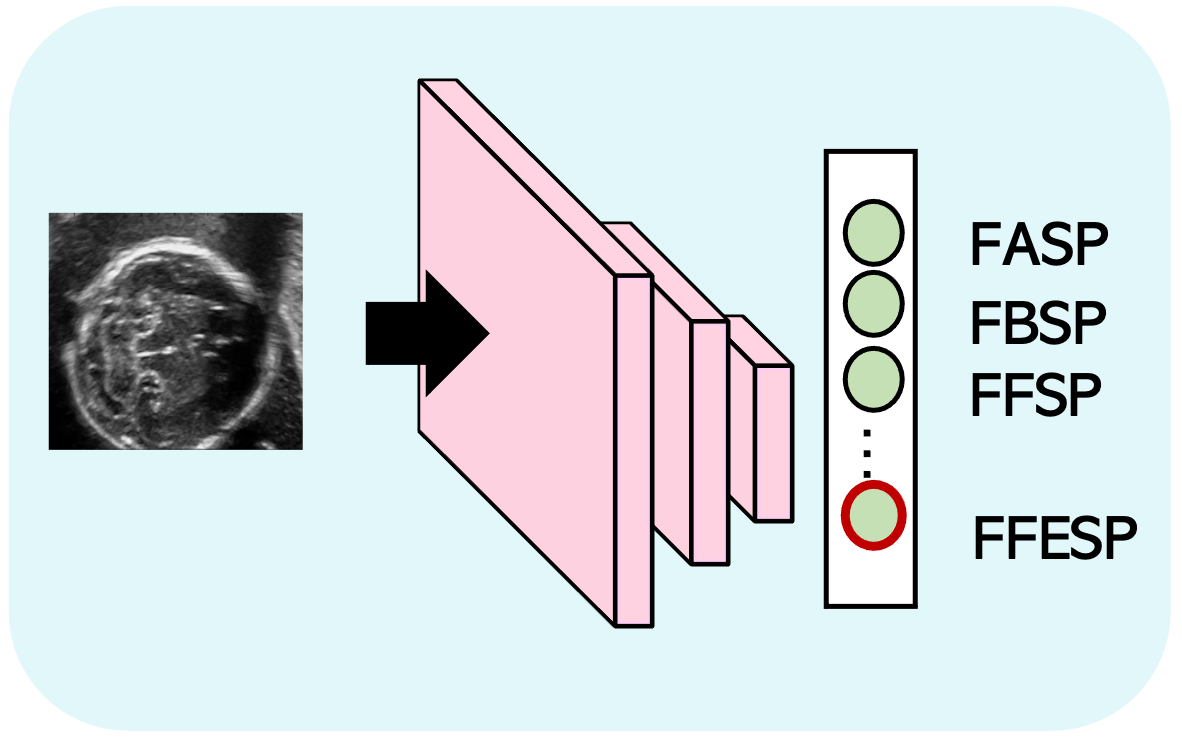}}
\subfigure[\label{subfig:classup}Classification + attention]
   {\includegraphics[width=.33\textwidth]{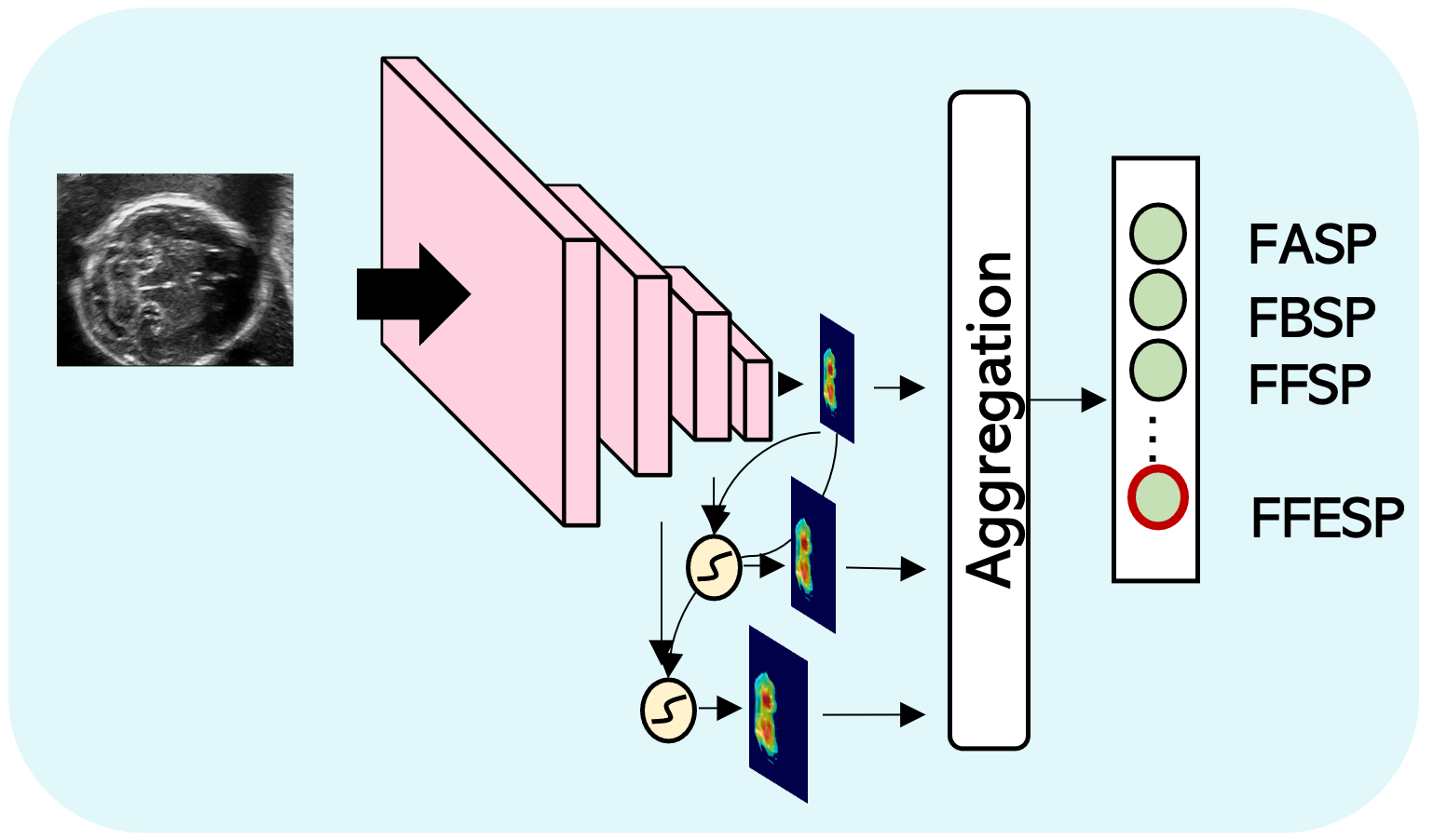}}
\subfigure[\label{subfig:det} Detection ]
   {\includegraphics[width=.33\textwidth]{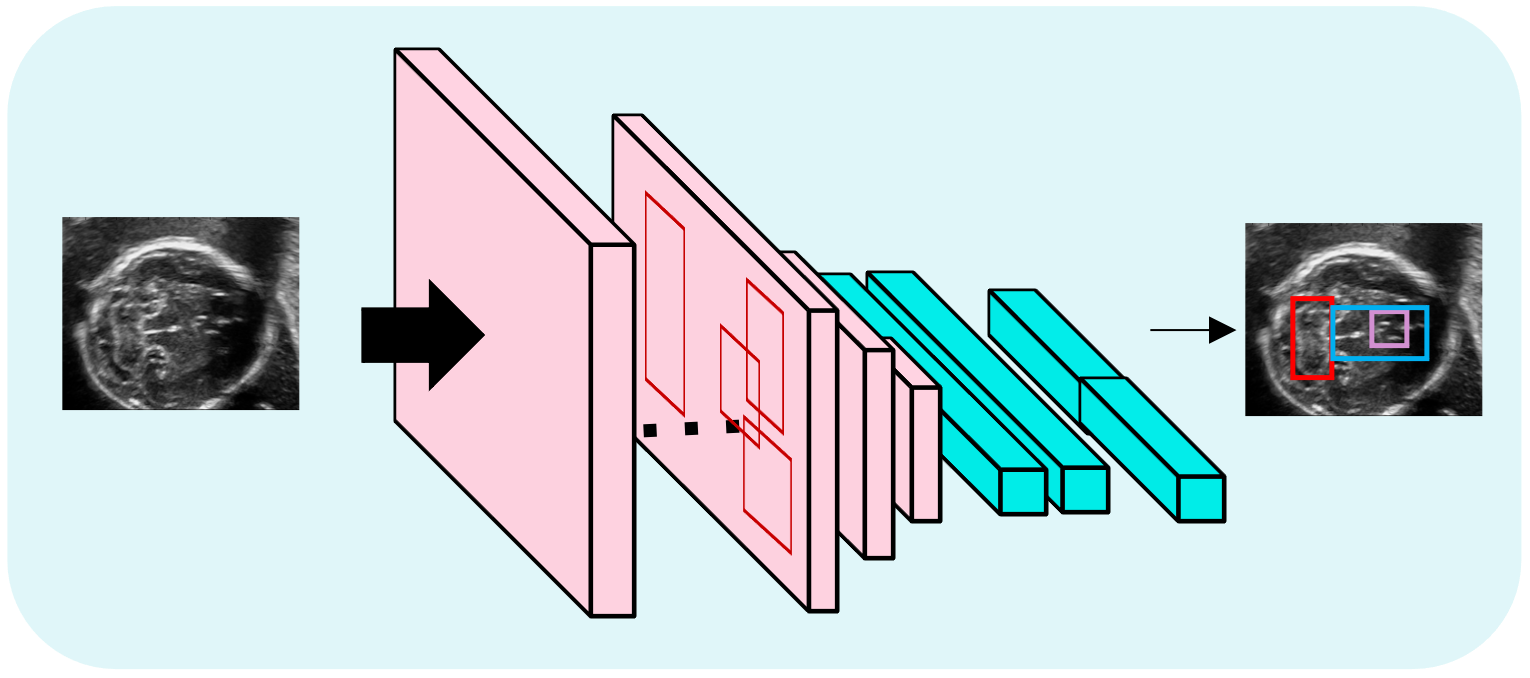}}
  \subfigure[\label{subfig:DRL} Deep reinforcement learning (DRL)]
   {\includegraphics[width=.35\textwidth]{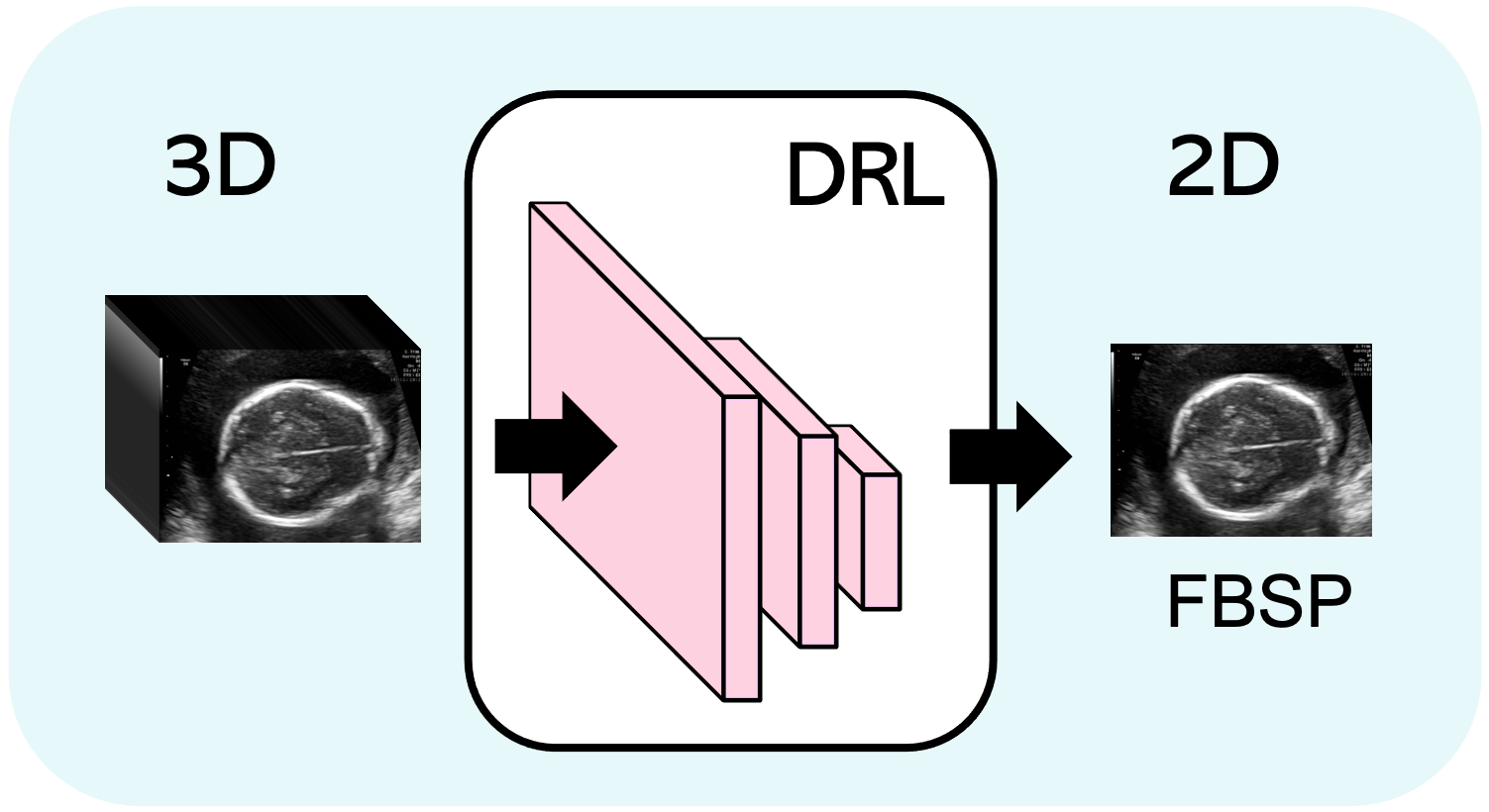}}
 \caption{\label{fig:overview_planes} 
 Most common deep-learning strategies to fetal standard-plane detection. FASP = fetal abdomen standard plane, FBSP = fetal brain standard plane, FFSP = fetal facial standard plane, FFESP = fetal femur standard plane. 
}
\end{figure}

In the clinical practice, the acquisition of a standard plane is performed manually by clinicians, which move the US probe across the mother's body until specific anatomical landmarks are visible in the image. 
Clinical expertise is required to face the high intra-class variability of US standard planes, which is due to, among the others, different gestational weeks, equipment vendors and US-probe angle \cite{pu2021automatic}. Moreover, the anatomical structures that characterize a specific plane could be common to other planes. 
DL algorithms may be a valuable tool to tackle these challenges.
Table~\ref{table:plane detection} summarizes the DL algorithms for standard-plane detection that are surveyed in this section. An overview of the most common strategies adopted in the literature is shown in Fig.~\ref{fig:overview_planes}.

%
One of the first DL algorithms for scan plane detection was proposed by \cite{wu2017fuiqa}, where the goal was to detect FASP. The algorithm consists of a first CNN that localizes fetal abdomen and a second one that detects the presence in the localized abdomen of stomach bubble and umbilical vein. The algorithm is tested on 2606 fetal abdominal images from 219 US videos, reaching a mean AUC of 0.99, along with mean $Acc$, $Rec$, $Spec$ of 0.98, 0.96 and 0.97, respectively. 
Despite the promising results, two separate CNNs are needed for localization and classification, respectively, inevitably increasing training and deployment time.
%


A number of algorithms tackles the problem of fetal standard plane detection as a mere classification problem. 
In \cite{yu2017deep}, FFSP is detected by fine tuning a shallow classification CNN pre-trained on ImageNet\footnote{https://www.image-net.org/}. The approach is tested on 2418 images reaching a mean AUC = 0.99 and $Acc$, $Prec$, $Rec$ and $F1$ of 0.96, 0.96, 0.97, 0.97, respectively. 
A shallow classification CNN is also used in \cite{qu2020standard} to automatically identify FBSPs. The dataset consists of 19142 images, which are augmented and split in 5 folds to test the algorithm. The CNN reaches $Acc$, $Prec$, $Rec$ and $F1$ of 0.93, 0.93, 0.92, 0.93, respectively. 
%
In~\cite{burgos2020evaluation}, state-of-the-art CNNs are compared to classify 6 planes (FASP, FTSP, FCSP, FVSP, and thorax standard planes). Testing is performed on a dataset of 5271 images (896 patients). The authors made the dataset available to foster research in the field (Sec. \ref{sec:dataset}, Table \ref{table:datasets}). The best performing CNN results to be DenseNet-169, with top-1 error, top-3 error and average class $Acc$ of 6.20\%, 0.27\% and 0.94, respectively.
A dense network is also used in \cite{kong2018automatic} to detect 4CH, FASP, FBSP, FFSPs. A total of 5678 US images are used as testing set, reaching $Prec$, $Rec$ and $F1$ of 0.98, 0.98 and 0.98, respectively.
Similarly, the work in \cite{liang2019sprnet} proposes an automatic fetal standard plane classification of 4CH, FASP, FBSP, FFASP, and coronal FFSP, based on DenseNet. The network is trained in conjunction with a placenta transferring dataset in order to discover and learn potential relationship between the dataset and possibly avoid overfitting. A test set of 4455 images is used, reaching $Acc$, $Rec$, $Spec$ and $F1$ of 0.99, 0.96, 0.99 and 0.95, respectively.
In \cite{sundaresan2017automated}, the detection of heart fetal standard planes (4CH, LVOT, 3VV and not heart) is addressed with a Fully Convolutional Network (FCN), that aims at segmenting the center of fetal heart and classifying the cardiac views in a single step. A total of 2178 frames is used as test set, reaching an error rate of 0.23.
A generative adversarial network (GAN) is used in \cite{montero2021generative} to improve FBSP classification by means of ResNet. A total of 2249 images is used to validate the approach, reaching an AUC of 0.86, along with $Acc$ and $F1$ of 0.81 and 0.80, respectively.
Cross-device classification of 6 anatomical standard planes (4CH, FASP, LVOT, RVOT, FFESP, Lips) is performed in \cite{meng2020unsupervised}. An improved feature alignment is used to extract discriminative and domain-invariant features across domains. 
The performance obtained in the target domain results in average in $F1$, $Rec$ and $Prec$ of 0.77, 0.77 and 0.78, respectively.

All these approaches work with 2D images. A couple of research papers extends the classification of standard planes to US video clips.
A DL framework is proposed in \cite{chen2017ultrasound} to detect FASP, FFASP and 4CH. To process the temporal information encoded in the US videos, the framework makes use of a long short-term memory (LSTM). For performance evaluation, 331 videos (13247 images in total), one from a different subject, are used. The mean $Acc$, $Prec$, $Rec$ and $F1$ in detecting the planes are 0.87, 0.71, 0.64, 0.64 respectively. 
Similarly, in~\cite{pu2021automatic}, a classification CNN and a recurrent neural network (RNN) are used to detect FASP, FTSP, FCSP, and FLVSP. A total of 224 videos consisting of 16740 frames in total is used to evaluated the method: mean $Acc$, $Prec$, $Rec$ and $F1$ of 0.85, 0.85, 0.85, 0.85 are obtained, respectively. 
%


To improve standard-plane detection performance and increase the interpretability of the DL results, attention mechanisms, which endow CNNs to better focus on the most discriminating regions in the US image, have been exploited.
The work in \cite{schlemper2019attention} is among the firsts to add a self-gated soft-attention mechanism to a CNN for the detection of 13 US standard planes. The framework is tested on 38243 images, reaching mean $Acc$, $Rec$, $Prec$ and $F1$ of 0.98, 0.93, 0.93 and 0.93, respectively. 
Another popular solution in the field is to include sonographer gaze attention mechanism. In \cite{cai2018multi}, a multi-task framework relying on SonoNet is proposed for FASP detection. The framework is trained to predict both standard plane and sonographer visual saliency prediction. To test the algorithm, 8 videos (324 frames in total), acquired from a different subject, are used. The inclusion of sonographer visual saliency prediction helps improving the standard-plane detection performance, achieving $Prec$, $Rec$ and $F1$ equal to 0.96, 0.96 and 0.96, respectively. 
In \cite{cai2020spatio}, a similar approach is proposed, which further processes US temporal clips via a temporal attention module. 
The considered planes are the FASP, FBSP and FFESP. A total of 280 videos, lasting from 3s to 7s, is split following five-fold cross-validation, reaching a mean $Prec$, $Rec$ and $F1$ of  0.89, 0.85 and 0.87, respectively.

Multiple data augmentation methods are exploited in \cite{lee2021principled} to improve standard plane classification. A three-fold cross-validation on 1129 standard plane frames (14 categories) and 1127 background images is used to verify the approach. A $Prec$ = 0.75, $Rec$ = 0.73 and $F1$ = 0.74 are obtained.

\setlength{\tabcolsep}{6pt} 
\renewcommand{\arraystretch}{1.5} 

\begin{table*}[tbp]
    
    \centering
    \caption{Summary of deep-learning algorithms for anatomical-structure analysis (for performance metrics refer to Table \ref{table:measures}). *: Only the total number of subjects is reported.}
    \label{tab:biometry}
    
    \begin{tabularx}{\textwidth}{XXXXlX}

        Paper (Year) & Organ & Training set size & Test set size &  Annotators & Performance metrics \\
        \hline 
        
        Dong et al., \cite{dong2019arvbnet} (2019) & Heart & - & 1991 2D & Few & $mAP$=0.93\\
        
        Patra et al., \cite{patra2019multi} (2019) & Heart & - (10 subjects) & - (2 subjects) & Few & $Acc$=0.82\\ 
            
        Huang et al., \cite{huang2017temporal} (2017) & Heart & 90 videos (12 subjects*) & 1 video (12 subjects*) & Few & - \\
        
        Petra et al., \cite{patra2017learning} (2017) & Heart & 89 videos (10 subjects) & 2 videos (2 subjects) & Few & $Acc$=0.79\\  
        
        Gao et al., \cite{gao2017detection} (2017) & Heart & 350 videos (412 subjects*) & 62 videos (412 subjects*) & 1 & ACC=0.90, $Prec$=0.85, $Rec$=0.89\\
        
        Nurmaini et al., \cite{nurmaini2020accurate} (2020) & Heart & 624 2D & 69 2D & 1 & Mean $DSC$=0.83\\
        
        Rachmatullah et al., \cite{rachmatullah2021convolutional} (2021) & Heart & 413 2D (3 subjects*) & 106 2D (3 subjects*) & Few & $IoU$=0.94, $Acc$=0.96\\
        
        Xu et al., \cite{xu2020dw} (2020) & Heart & 716 2D (895 subjects*) & 179 2D (895 subjects*) & Few & $DSC$=0.83, $Acc$=0.93, AUC=0.99\\
        
        Xu et al., \cite{xu2020convolutional} (2020) & Heart & 1284 2D & 428 2D & Few & $DSC$=0.85, $HD$=3.33, $Acc$=0.93\\
        
        Yu et al., \cite{7744576} (2017) & Heart & 41 videos (40 frames each) & 10 videos (40 frames each) & 1 & $HD$=1.26, $AD$=0.20, $DSC$=0.94\\
        
        An et al., \cite{an2021category} (2021) & Heart & 512 2D (319 subjects*) & 126 2D (319 subject*) & 10 & $DSC$ = 0.78, $AP$ = 0.45 \\
        
        Nurmaini et al., \cite{nurmaini2021deep} (2021) & Heart & 1033 2D (50 subjects*)  & 116 2D (50 subjects*) & 1 & $mAP$=0.96, $IoU$=0.79, $DSC$=0.90\\
        
        Tan et al., \cite{tan2020automated} (2020) & Heart & - (100 subjects*) & - (100 subjects*) & Few & $Prec$=0.85, $Rec$=0.86, $F1$=0.86, AUC=0.93\\
        
        Dozen et al., \cite{dozen2020image} (2020) & Heart & 410 2D (211 subjects*) & 205 2D (211 subjects*) & Few & $IoU$=0.55, $DSC$=0.68\\
        
        Komatsu et al., \cite{komatsu2021detection} (2021) & Heart & 213 videos (363 subjects*) & 34 videos (363 subjects*) & Few & $mAP$=0.70, AUC=0.83\\
        
        Arnaout et al., \cite{arnaout2021ensemble} (2021) & Heart & 107823 2D (1326 subjects) & 4867591 2D (4666 subjects) & Few & AUC=0.94, $Rec$=0.95, $Spec$=0.96\\
        
        Qiao et al., \cite{qiao2022rlds} (2022) & Heart & 2000 2D (2000 subjects) & 100 2D (100 subjects) & 2 & $Prec$=0.93, $Rec$=0.93\\
        
        Gong et al., \cite{gong2019fetal} (2019) & Heart & 3596 2D & 400 2D & Few & $Acc$ = 0.85\\
        
        Pu et al., \cite{pu2021fetal} (2021) & Heart & 586 videos (minimum 80 and maximum 373 frames) (350 subjects*) & 151 videos (minimum 80 and maximum 373 frames) (350 subjects*) & Few & $Acc$=0.95, $Rec$=0.93, $Spec$=0.94, $F1$=0.95\\
        
        Patra et al., \cite{patra2020hierarchical} (2020) & Heart & 89 videos (39556 2D) (12 subjects) & 2 videos (12 subjects) & Few & -\\ 

        
        Wang et al., \cite{Wang_2018} (2018) & Brain & - & 4005 2D (1783 subjects) & Few & $DSC$=0.77, $IoU$=0.63, $HD$=26.40 pixel \\
        
        Wu et al., \cite{Wu_2020} (2020) & Brain & - & 448 2D (224 subjects) & 3 & $Prec$=0.79, $Rec$=0.74, $DSC$=0.77, $HD$=0.78\\
        
        Singh et al., \cite{Singh_2021} (2021) & Brain & 588 2D & 146 2D & Few & $DSC$=0.87, $HD$=28.15, $Rec$=0.86, $Prec$=0.90\\
        
        Zhang et al., \cite{zhang2020multiple} (2020) & Brain & 718 2D (HC18) & 70 2D (HC18) & 2 & $DSC$=0.97, $Prec$=0.97, $Rec$=0.98, $HD$=10.92 mm\\
        
        Huang et al., \cite{huang2018vp} (2018) & Brain & 200 3D & 45 3D & Few & $IoU$=0.63\\
        
        Wyburd et al., \cite{Wyburd_2020cortical} (2020) & Brain & 271 3D & 36 3D & Few & $DSC$=0.82, BCE=0.09\\

    \end{tabularx}
\end{table*}

\setcounter{table}{4}
\begin{table*}[tbp]
    
    \centering
    \label{tab:biometry}
    
    \begin{tabularx}{\textwidth}{XXXXlX}

        Paper (Year) & Organ & Training set size & Test set size & Annotators & Performance metrics \\
        \hline
        
        Venturini et al., \cite{Venturini_2020} (2020) & Brain & 480 3D & 48 3D & Few & $DSC$=0.83, $ED$=2.24 mm, $HD$=4.51 mm\\
        
        Hesse et al., \cite{hesse2020improving} (2020) & Brain & 138 3D & 15 3D & Few & Surface Dice=0.92, Average Surface Distance=0.23 mm\\
        
        Yang et al., \cite{yang2020hybrid} (2020) & Brain & 50 3D (100 subjects*) & 50 3D (100 subjects*) & 2 & $DSC$=0.96, $IoU$=0.92, $HD$=0.46 mm\\
        
        Namburete et al., \cite{namburete2018fully} (2018) & Brain & Not specified 2D (from 599 3D) & Not specified 2D (from 140 3D) & Few & $Acc$=0.99, $ED$=6.9 mm, $IoU$=0.82, $HD$=9.3 mm\\
        
        Moser et al., \cite{moser2020automated} (2020) & Brain & 885 3D & 300 3D & Few & $ED$=1.36 mm, $HD$=9.05 mm, $DSC$=0.94\\
        
        Xie et al., \cite{Xie_2020} (2020) & Brain & 29419 2D (12780 subjects*) & 4739 2D (12780 subjects*) & 15 & $mAP$=0.98, $Rec$=0.94, $DSC$=0.94, $Acc$=0.96, $Spec$=0.96, AUC=0.99\\ 
        
        Lee et al., \cite{lee2020calibrated} (2020) & Brain & 8369 2D & 869 2D & Few & $RMSE$=13.85, $MAE$=16.05\\
        
        Namburete et al., \cite{Namburete2017Robust} (2017) & Brain & 326 3D & 121 3D & Few & Prediction error=6.9\\
        
        Wyburd et al., \cite{wyburd2021assessment} (2021) & Brain & 689 3D & 122 3D & Few & $MAE$=4.1, for Sylvian fissure; $MAE$=5.1, for Parieto-occipital fissure; $MAE$=4.9, for Calcarine fissure\\

        Hu et al., \cite{hu2019automated} (2019) & Placenta & 1363 2D (247 subjects*) & 205 2D (247 subjects*) & 3 & $DSC$=0.92, $Acc$=0.93\\
        
        Hu et al., \cite{hu2021automatic} (2021) & Placenta & 10707 2D (321 subjects*) & 2677 2D images (321 subjects*) & - & $Acc$=0.81, $Rec$=0.88, $Spec$=0.65\\
        
        Zimmer et al., \cite{zimmer2020multi} (2020) & Placenta & 27081 2D (67 subjects*) & 5149 2D (67 subjects*) & 1 & $DSC$=0.82, $HD$=31.84 mm\\
        
        Oguz et al., \cite{oguz2018combining} (2018) & Placenta & - (47 subjects*) & - (47 subjects*) & 2 & $DSC$=0.86\\
        
        Looney et al., \cite{looney2018fully} (2018) & Placenta & 1196 3D (3104 subjects) & 1197 3D (3104 subjects) & 3 & $DSC$=0.84, $HD$=14.6 mm\\ 
        
        Looney et al., \cite{looney2017automatic} (2017) & Placenta & 280 3D (3064 subjects*) & 20 3D (3064 subjects*) & Few & $DSC$=0.73, $HD$=27 mm\\
        
        Zimmer et al., \cite{zimmer2019towards} (2019) & Placenta & 30 3D (30 subjects*) & 12 3D (30 subjects*) & - & $DSC$=0.80\\
        
        Yang et al., \cite{yang2017towards} (2017) & Placenta & 60 3D (104 subjects*) & 44 3D (104 subjects*) & 10 & $DSC$=0.64, $HD$=24.54 mm\\
        
        Li et al., \cite{li2017automatic} (2017) & Amniotic Fluid & 400 2D & 900 2D & Few &  $Acc$=0.78, $IoU$=0.54\\
        
        Cho et al., \cite{cho2021automated} (2021) & Amniotic Fluid & 310 2D (100 subjects*) & 125 2D (155 subjects*) & 1 & $DSC$=0.87, $Prec$=0.89, $Rec$=0.87, $Spec$=0.99\\
        
        Sun et al., \cite{sun2021complementary} (2021) & Amniotic Fluid & 2200 2D (1190 subjects*) & 180 2D (1190 subjects*) & 3 & $DSC$=0.86, $Rec$=0.81, $Prec$=0.93, $HD$=15.49 mm\\ 

        
         
        Xia et al., \cite{Xia_2021} (2021) & Lung & 6312 2D & 701 2D & - & $Acc$=0.83, $Rec$=0.82, $Spec$=0.83, AUC=0.95 \\
        
        Chen et al., \cite{chen2020preliminary} (2020) & Lung & 206 2D & 126 2D & 1 & $MAE$= 1.56 mm \\
        
        Weerasinghe et al., \cite{Weerasinghe_2020} (2020) & Kidney & 40 3D & 60 3D & 2 & $DSC$=0.81, $IoU$=0.69, $HD$=8.96 mm \\
        
        Franz et al., \cite{franz2021deep} (2021) & Spine & 320 3D & 80 3D & - & - \\  
        
        Chen et al., \cite{chen2021neural} (2021) & Spine & - & - & 2 & $Rec$=0.93, $Prec$=0.96, $Acc$=0.94, $ IoU$=0.91 \\ 
        
        Schmidt-Richberg et al., \cite{ schmidt2017abdomen} (2017) & Abdomen & 126 3D & 42 3D & - & $MAE$=2.24 mm\\
        
        Droste et al., \cite{droste2019towards} (2019) & Abdomen & 30 videos & 3 videos & Few &  AUC by Judd=0.87, $D_{KL}$= 2.16\\

    \end{tabularx}
\end{table*}

\setcounter{table}{4}
\begin{table*}[tbp]
    
    \centering
    \label{tab:biometry}
    
    \begin{tabularx}{\textwidth}{XXXXlX}

        Paper (Year) & Organ & Training set size & Test set size &  Annotators & Performance metrics \\
        \hline 
        
        Wang et al. \cite{wang2019joint} (2019) & Femur & 30 3D (30 subjects) & 20 3D (20 subjects) & 1 & $DSC$=0.91, $IoU$=0.83, $HD$=4.08 mm, $ED$=0.87 mm\\
        
        Cerrolaza et al., \cite{cerrolaza2018deep} (2018) & Head & 52 3D & 16 3D & Few & $DSC$=0.83, $IoU$=0.70\\
        
        Perez-Gonzalez et al., \cite{perez2020deep} (2020) & Head & 10 3D (18 subjects*) & 8 3D (18 subjects*) & 1 & $DSC$=0.81, AUC=0.82\\
        
        Wang et al., \cite{wang2020annotation} (2020) & Head & 799 2D (HC18) & 200 2D (HC18) & Few & $DSC$=0.94\\
        
        Sridar et al., \cite{sridar2019decision} (2019) & Multiple & 3109 2D & 965 2D & 1 & $Acc$=0.97, $Prec$=0.76, $Rec$=0.75\\
        
        Ishikawa et al., \cite{ishikawa2019detecting} (2019) & Body, Head, Leg & 10000 2D & 2000 2D & 1 & $Rec$=0.92\% \\ 
        
        Sharma et al., \cite{sharma2019spatio} (2019) & Multiple & 31629 2D & 3986 2D & Few & Top-1 $ACC$=0.77, Top-3 $ACC$=0.94 \\
        
        Chen et al., \cite{chen2020cross} (2020) & Abdomen, Heart, Skull & 225825 (free hand) and 30048 (single sweep) 2D   & 7512 (single sweep) 2D & - & $Acc$=0.73 and $Acc$=0.89 for 3- and 4-class classification, respectively \\ 
        
        Alsharid et al., \cite{alsharid2020curriculum} (2020) & Abdomen, Head, Heart, Spine & 12808 2D & 9979 2D & - & $Prec$=0.96, $Rec$=0.96, $DSC$=0.96 \\
        
        Xu et al., \cite{xu2018less} (2018) & Kidney, Liver, Spleen & 149775 2D & 37444 2D & 2 & $Acc$=0.85 \\
        
        Gao et al., \cite{gao2019learning} (2019) & Abdomen, Heart, Skull & 365 videos & 91 videos & - & $AP$=0.91 \\ 
        
        Wu et al., \cite{wu2017cascaded} (2017) & Abdomen, Head & 1588 2D & 741 2D & 1 & $DSC$= 0.97, $IoU$=0.96\\
        
        Yang et al., \cite{Yang2019TowardsAS} (2019) & Fetus, Gestational sac, Placenta & 60 3D & 44 3D & 10 & $DSC$=0.80, $HD$= 14.13 mm \\
        
    \end{tabularx}
\end{table*}


Using classification CNNs may lead to inaccurate detection of standard planes since it does not involve the detection of any specific anatomical landmarks. This, in fact, does not reflect the way clinicians detect standard planes.
A different approach investigated in the literature is to accomplish standard-plane detection through  anatomical-structure detection. 
This approach is followed in \cite{dong2019generic}, where a classification CNN identifies images of the 4CH and a multi-task SSD detects the key anatomical structures of the plane as well as the US gain parameter and zoom of the image. The algorithm is validated by means of a five-fold cross-validation on 7032 images. The authors obtain a $mAP$ of 0.81.
A Faster R-CNN is used to assess the presence of fetal head US images' specific anatomical structures in \cite{lin2018quality}. A total of 1153 images is used for testing, achieving $AP$, $Rec$ and $Prec$ of 0.79, 0.85, 0.87, respectively in detecting the structures of interest.
A similar approach is followed in~\cite{lin2019multi}. A multi-task framework, consisting of a Faster R-CNN  with an additional classification branch, is used to detect 6 key anatomical structures and evaluate if the head is centered in the image. A $mAP$ of 0.93 is obtained on 320 test images.
A multi-task framework is also used in~\cite{zhang2021automatic}, in which a CNN inspired by Faster R-CNN is used to both predict the presence of specific anatomical structures in abdomen, brain and heart images and classify these structures. A total of 820 images is used to test the framework, achieving a $mAP$ of 0.95 in detecting structures and $Acc$, $Prec$ and $Rec$ of 0.95, 0.95, 0.93, respectively in classify them. 
A hybrid approach is proposed in \cite{baumgartner2017sononet}, where a CNN is trained to localise 13 anatomical structures by means of a weak supervision provided by image-level labels, thus, without the need for bounding-box annotation during training. The approach is tested on 109165 images and 200 videos. Mean $Prec$, $Rec$ $F1$ and $IoU$ of 0.77, 0.90, 0.80 and 0.62 are achieved, respectively.
In \cite{yaqub2017deep},  FVSP detection is accomplished by localizing fetal brain by means of a segmentation CNN at first. Then, CSP visibility, fetal brain symmetry, and midline orientation are assessed by means of a number of additional CNNs. A five-fold cross-validation on 19838 images is used to evaluated the framework, reaching an achievable FVSP detection in more than 95\% of cases.

More recently, semi-supervised and self-supervised strategies for standard-plane detection have been investigated.
In~\cite{gao2020label}, a semi-supervised pipeline is proposed to detect FTSP and FCSP from freehand fetal US video. The framework consists of a CNN for feature extractor, a prototypical learning module and a semantic transfer module to automatically label unseen video frames. A total of 60 videos (20 subjects) is used as test set, reaching a $mAP$ of 0.87.
In~\cite{chen2019self}, a self-supervised method for scan plane detection in fetal 2D US images is proposed. Specifically, given an image, two small patches are randomly selected and swapped on the image and this procedure is repeated multiple times. A CNN is trained to restore the altered image back to its original version. The CNN weights are then used to perform classification of fetal standard planes (FASP, FBSP, FFESP, Kidney, Spine, 4CH, 3VV, RVOT, LVOT and facial profile. When tested on the same dataset as the one used in~\cite{baumgartner2017sononet}, the method achieves $Prec$, $Rec$ and $F1$ of 0.89, 0.90 and 0.89, respectively.
A semi-supervised learning approach to classify 13 standard planes is exploited in \cite{tan2019semi}. 100 images for each class from a dataset of 22757 images are used as labelled data. The remaining images are treated as unlabelled data. 
For a test set of 5737 images, overall accuracy is close to 0.70. 


A number of researchers is working to perform standard plane detection using DRL from 3D fetal US, mainly to mimic what clinicians do and exploring inter-plane dependency. 
The work in \cite{dou2019agent} proposes a DRL to localize fetal brain standard planes in US volumes. The DRL framework is equipped with a landmark-aware alignment module that exploits a CNN to detect anatomical landmarks in the US volume. The landmarks are then registered to a plane-specific atlas. The DRL agent's interaction procedure is terminated by means of a RNN module. The efficacy of the method is validated on 100 US volumes, obtaining an average theta of 9.36\degree along with a mean $DF$ of 3.03 mm are obtained. 
The approach is further enhanced in~\cite{yang2021agent}, which designs an adaptive RNN-based termination module to early stop the agent searching. To validate the method, 100, 110 and 144 volumes of fetal brain, fetal abdomen and uterus, respectively, are used. 
A mean $DF$ of 2.31 mm along with a mean $\theta$ of 10.36\degree are obtained. 
A similar approach is performed in~\cite{yang2021searching}, which localizes multiple uterine standard planes in 3D simultaneously by multi-agent DRL. The latter is equipped by one-shot neural architecture search (NAS) module. To improve system robustness against the noisy environment, a landmark-aware alignment model is utilized. The spatial relationship among standard planes is learnt by a RNN. The method is evaluated by means of 144 volumes of fetal brain and uterus. A mean $\theta$ and $DF$ of 7.20\degree and  1.82 mm are obtained, respectively. 

US volumes are also processed in \cite{li2018standard}, where 2D image planes are fed to a CNN that predicts the 3D transformation to register each plane to a ground truth standard plane. The process is evaluated on 3D US volumes of fetal brain from 22 subjects. A mean plane centre difference of 3.44 mm along with rotation angle between the planes of 11.05\degree  are obtained. 

\subsection{Limitations and open issues for fetal standard-plane detection}


From the review of the literature on standard-plane detection,  a first limitation emerges in the number of considered planes.
There are papers (\cite{chen2017ultrasound,yu2017deep,qu2020standard,pu2021automatic}) that consider only few planes. In some cases (\cite{cai2018multi,cai2020spatio,wu2017fuiqa}) only one plane is considered. 

When focusing on the experimental setup, only few papers refer to US scans on patient basis. This information is, however, of crucial importance since US scans belonging to a woman should not be part of both training/validation and test set to avoid bias. %
A second issue arises in terms of quantitative comparison among the different approaches. The use of different datasets, some of which are really small in size, does not allow to make a fair comparison among the solutions proposed in the literature. 

An important step towards a fair evaluation of methods has been done in \cite{burgos2020evaluation}, which released the first dataset in the field. However, only FASP, FBSP, 4CH, maternal cervix, FFESP are considered.
Showing training/validation curves as well as using visual explanation techniques (i.e. Grad-CAM \cite{selvaraju2017grad}) should also be considered for a fair assessment of algorithm performance, also in terms of model bias and variance.


\section{Anatomical-structure analysis}
\label{sec:organs}

\begin{figure}[tbp]
    \centering
    \includegraphics[width = .5\textwidth]{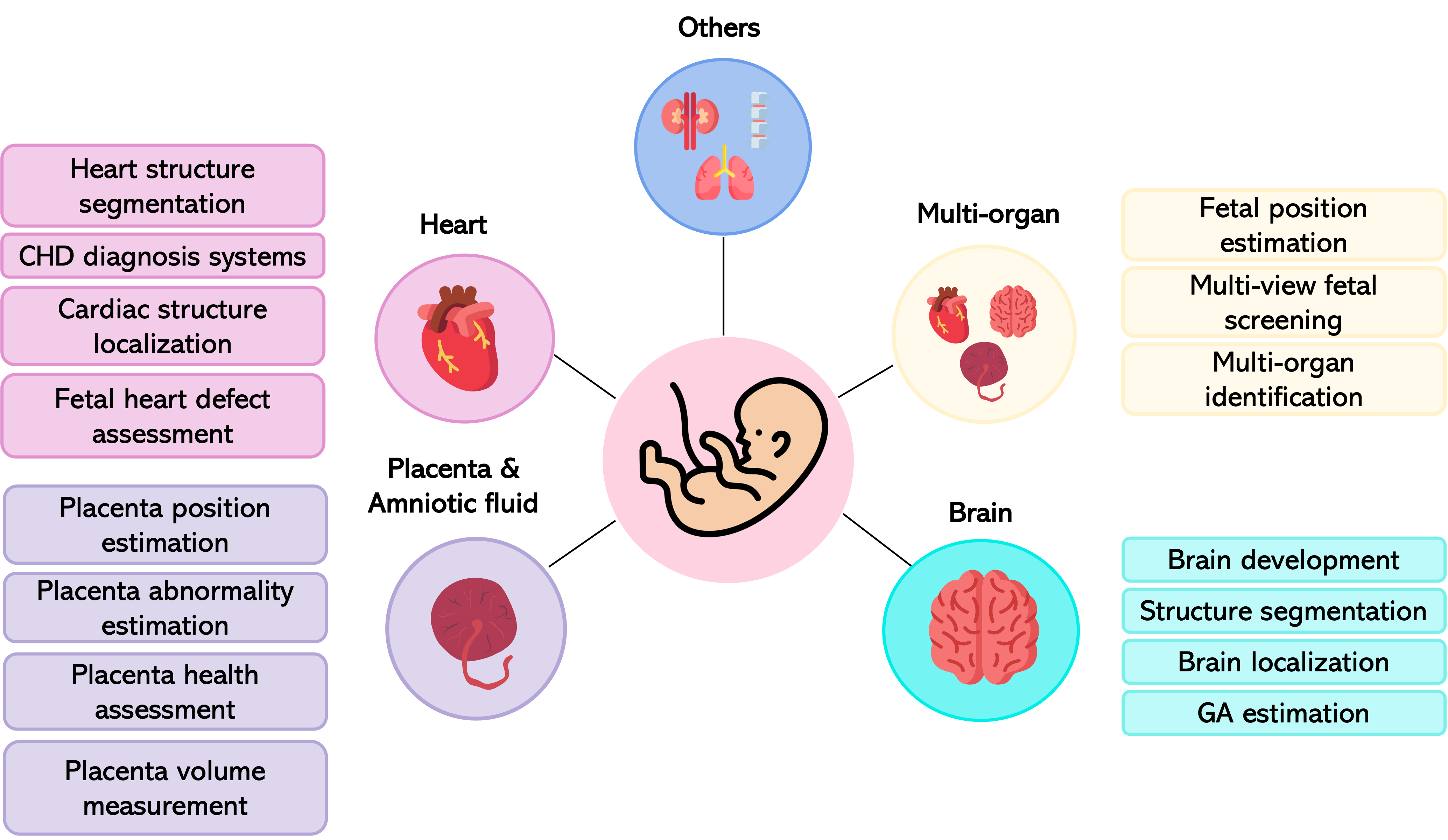}
  \caption{Overview of the most common tasks in the literature for fetal organs analysis.}
    \label{fig:overview_structures}
\end{figure}

This section surveys methods for the analysis of fetal heart (Sec. \ref{sec:heart}), brain (Sec. \ref{sec:brain}), and placenta and amniotic fluid (Sec. \ref{sec:placenta}), for which a relatively rich literature already exists.  DL approaches for the analysis of other fetal anatomical-structures are grouped in Sec.~\ref{sec:other_organs}, while Sec.~\ref{sec:multiple_organs} surveys approaches for multi-structure analysis. 
Figure~\ref{fig:overview_structures} summarizes the main tasks addressed in the literature of anatomical-structure analysis.

\subsection{Heart} \label{sec:heart}
Fetal cardiac evaluation is of crucial importance to detect heart diseases, such as congenital heart diseases (CHDs), and intrauterine growth restriction. Cardiac evaluation generally consists of cardiac-function analysis and heart anatomical evaluation, including heart dimension and shape.

A number of DL approaches in the field focuses on heart and heart-structure detection. 
The work in \cite{dong2019arvbnet} uses an SSD model with aggregated residual visual blocks to detect  anatomical-heart structures such as left atrial pulmonary vein angle, apex cordis, moderator band, and multiple ribs in 4CH. Experiments on 1991 4CH show a $mAP$ = 0.93.
A cardiac-structure localization algorithm is proposed in \cite{patra2019multi}. 
The presence of the heart in the 4CH is detected using a modified VGG-16, then a Faster RCNN model coupled with LSTM layers is used to temporally classify the presence of Foramen ovale, Mitral valve, Tricuspid valve, LV wall and RV wall. Two out of 12  subjects’ videos, which are in total 91 (39556 frames), are  used as test set.
An $Acc$ = 0.82 is reached.
In \cite{huang2017temporal}, the presence, viewing plane, location and orientation of the fetal heart is predicted by means of a recurrent CNN. A total of 91 fetal cardiac screening videos from 12 subjects is annotated at frame level. A leave-one-out cross-validation over each subject, resulting in 12-fold validations, is used to evaluate the accuracy of  classification and localization. A custom based metric is proposed, which takes classification and localization results into account. 
In \cite{patra2017learning}, the same task as in \cite{huang2017temporal} is formulated as a multi-task learning problem within a hierarchical convolutional model that progressively encodes temporal information throughout the network. The dataset consists of 91 videos from 12 subjects, 2 of which are used as test. An $Acc$ of 0.83 in correctly classifying views and an $Acc$ = 0.79 in correctly localize structures are reached.
In \cite{gao2017detection}, spatio-temporal representations of fetal heart are learned by means of an end-to-end two-stream fully CNN for temporal sequence analysis. The goal is to captured motion and appearance features in a weakly supervised manner. The 15\% of 412 fetal US videos is used as test. In terms of heart identification, a 0.90, 0.85 and 0.89 of ACC, $Prec$ and $Rec$, respectively are obtained. 

Structure segmentation offers more information than detection, since heart and heart-structure shape is a significant indicator of possible pathology. Here, encoder-decoder CNNs are often exploited in the literature.
In \cite{rachmatullah2021convolutional}, a U-Net architecture is used to segment fetal cardiac standard planes to early detect possible structural heart abnormalities. The testing data consists of 106 images including atrial septal defects and ventricular septal defects. An $IoU$ and $Acc$ of 0.94 and 0.96 are obtained, respectively.
A cascaded network is developed in \cite{xu2020dw} to accurately segment 7 anatomical structures in the 4CH view. The network consists of a dilated sub-network responsible for aggregating both global and local information and two stacked U-Nets. A five-fold cross-validation on 895 4CH views (895 health women) is used to validate the approach. Mean $DSC$, $Acc$ and AUC of 0.83, 0.93, 0.99 are obtained, respectively. 
A cascaded U-Net is also used in \cite{xu2020convolutional} to segment anatomical heart structures, reaching $DSC$, $HD$ and  $Acc$ of 0.85, 3.33, 0.93, respectively on 428 images.
Segmentation of fetal left ventricle from fetal echocardiac videos is proposed in~\cite{7744576}. A dynamic CNN trained with multiscale information and different fine tuning strategies is used. The first cardiac frame is labelled to fine-tune the CNN.  As the frames are segmented sequentially, the CNN is fine-tuned dynamically by shallow tuning to fix the latest frame. To differentiate the left ventricle and atrium regions, the mitral valve base points were tracked. 41 sequences (40 frames each) are used as testing set. The CNN achieves $HD$, $AD$, $DSC$ of 1.26, 0.20, 0.94, respectively.

Instance segmentation of heart structures is performed in a number of papers, including
 \cite{an2021category} where the four cardiac chambers are segmented with a  network with three branches: the category branch, the mask branch, and a category-attention branch. This latter is used to correct the instance mis-classification and improve the segmentation accuracy. A total of 638 images from 319 fetuses is used for testing, following a five-fold cross-validation. Mean $DSC$ = 0.79, 0.76, 0.82, 0.75 for the four cardiac chambers, respectively, is obtained, with $mAP$ of 0.45. 
A Mask-RCNN is used in \cite{nurmaini2020accurate} to detect and segment the left  (LA) and right atrium (RA), left (LV) and right ventricle (RV), aorta and hole. The proposed approach is used to assess fetal heart defects (atrial septal, ventricular septal and atrioventricular septal defects). The 10 \% of 693 images is used as testing set. A $DSC$ = 0.84 (aorta), = 0.68 (hole), = 0.88 (LA), 0.89 (RA) and = 0.87 are obtained. 
A Mask RCNN is used in \cite{nurmaini2021deep} to segment the fetal heart in 4 different standard views (3VT, 4CH, LVOT, RVOT) and fetal heart chambers in each view to search for possible heart defects. A total of 116 US images is used as testing set. A $mAP$ = 0.96, $IoU$ = 0.79 and $DSC$ = 0.90 are reached in the detection and segmentation of standard view fetal heart. A $IoU$ = 0.72 in detecting heart structures (e.g. left ventricle, right atrium etc) in fetal standard views is obtained. 

CHD diagnosis is another popular task investigated in the literature of fetal heart analysis.
A hypoplastic left heart syndrome (HLHS) detector is developed in \cite{tan2020automated}. The detector consists of a SonoNet to detect standard planes (4CH, LVOT, RVOT) and a VGG16 to identify HLHS patients versus healthy subjects. Starting from US images from 100 fetuses, data curation and split is performed manually. Test set size in terms of fetuses and frames is not specified. The authors achieve $Prec$, $Rec$, $F1$ and AUC of 0.85, 0.86, 0.86 and 0.93, respectively.
In \cite{dozen2020image}, ventricular septal defects are evaluated by means of YOLOv2, which detects the ventricular septum, and U-Net that is used to segment the cropped ventricular septum area. A calibration module is used to further enhance U-Net output. A total of 615 images (421 videos) is extracted from 211 fetuses. A three-fold cross-validation is used for validation. A $IoU$ and $DSC$ of 0.55 and 0.68 are obtained, respectively.
YOLOv2 is also used in \cite{komatsu2021detection} for the detection of abnormalities in cardiac cardiac from 4CH and 3VT videos. The analysis is performed on 2D frames extracted from 34 videos using three-fold cross-validation. A $mAP$ of 0.70 is achieved in detecting cardiac substructures along with a mean AUC of 0.83 in the assessing of cardiac structural abnormalities detection. 
In \cite{arnaout2021ensemble}, an ensemble of deep residual networks are used to classify five heart views and successively classify normal hearts and complex CHD. A modified U-Net is further trained to calculate cardiothoracic parameters such as the cardiothoracic ratio (CTR), cardiac axis (CA) and fractional area change (FAC) for each cardiac chamber. Four datasets (FETAL-125: 19822 images, OB-125: 329405 images, OB-4000: 4473852 images, BCH-400: 44512 images) are used for testing. The approach achieves a mean AUC = 0.94 in classifying normal and abnormal hearts. 
In \cite{qiao2022rlds}, a CNN with a residual learning module linked with guided back-propagation to visualize feature maps is used to diagnose fetal CHD. The framework is evaluated by means of 100 4CH views (50 healthy fetal heart and 50 CHD). Framework reaches a $Prec$ = 0.93 and $Rec$ = 0.93. 
In \cite{gong2019fetal}, a one-class classification network is proposed to classify patients with CHD and healthy subjects. An improved GAN is used for data augmentation. A balanced dataset of 400 testing images is used as testing set to compared the the approach with state-of-the art ones. An $Acc$ = 0.85 is obtained. 
In \cite{pu2021fetal}, a YOLOv3 is exploited to detect the four-chamber views at first. Then a mobileNet is used to classify samples into ED or ES. A total of 151 videos is used to evaluate the framework. The method achieves $Acc$  $Rec$, $Spec$ and $F1$ of 0.95, 0.93, 0.94, 0.95, respectively. 
In order to tackle lack of available data in fetal CHD domain, the work in \cite{patra2020hierarchical} proposes an incremental learning approach to build a hierarchical network model that allows for a parallel inclusion of previously unseen anatomical classes without requiring prior data distributions. The goal here is the detection of different anatomical structure in different fetal cardiac views. The pipeline relies on natural hierarchies in US videos and it is built to account for new data in a self-organized fashion. Two out of 12 subjects' videos, which are in total 91 (39556 frames), are used as test set. Results, presented with custom metrics, highlight the benefits of incremental learning.


\subsection{Brain} \label{sec:brain}

Assessing fetus brain development is crucial to evaluate fetus growth and diagnose brain pathologies. In the clinical practice, several structures are analyzed to assess the fetal brain development and well-being.
From the clinician side, this is not effortless due to challenges such as the high intra- and inter-structure variability. 
Automatic fetal-brain analysis commonly includes anatomical-structure localization, segmentation, classification, and measurement. Structures considered in the literature include middle cerebral artery, cavum septum pellucidum (CSP), cerebellum, brainstem, ventricles, thalamus, cortical plates and fissures (Sylvian, Parieto-occipital, calcarine).
%


Among DL approaches for fetal brain analysis, a number of work focuses on brain-structure segmentation using encoder-decoder architectures. 
The work in~\cite{Wang_2018} proposes a DL-based architecture to segment the middle cerebral artery and provide the gate position on Doppler US images to the sonographer. The architecture uses a pre-trained dilated residual network as encoder and dense upsampling convolution blocks as decoder. After the segmentation, the gate position is retrieved as the center of the segmented area. When training the model on a dataset of 4005 US images, the authors achieve $DSC$, $IoU$ and $HD$ of 0.77,  0.63, and 26.40 pixel, respectively.  
The work in~\cite{Wu_2020} proposes a deep attention network inspired by the U-Net encoder-decoder architecture to segment the CSP and measure its width to evaluate the presence of anomalies. VGG11 is used as backbone in the encoding path and a channel attention module is introduced between the encoder and the decoder. Segmentation results obtained on a dataset of 448 US images acquired both on trans-thalamic axial and sagittal planes are: $Prec$ = 0.79, $Rec$ = 0.74, $DSC$ = 0.77, $HD$ = 0.78. 
Similarly, a U-Net inspired CNN (i.e., ResU-Net) is used in~\cite{Singh_2021}, with the aim to automatically segment the cerebellum. ResU-Net encodes residual blocks and dilated convolutions to recover the spatial resolution lost in the encoder keeping the number of training parameters low. A total of 734 US images acquired in the fetal trans-cerebellar plane is used following five-fold cross-validation, achieving a mean $DSC$, $HD$, $Rec$ and $Prec$ of 0.87, 28.15, 0.86, and 0.90 respectively. 
In \cite{zhang2020multiple}, a U-Net inspired CNN (i.e., MA-Net) is proposed for fetal head circumference segmentation. MA-Net is based on an encoder-decoder architecture consisting of 5 modules: encoder, atrous convolution, pyramid pooling, decoder and residual skip pathway modules. Results obtained testing the model on 70 images of the HC18 dataset are: $DSC$ = 0.97, $Prec$ = 0.97, $Rec$ = 0.98 and $HD$ = 10.92 mm.

More recently, a number of segmentation approaches has been exploring 3D architectures for processing the 3D information naturally encoded in 3D US data. 
A 3D U-Net is used in \cite{huang2018vp} to mask out extra-cranial tissues as a preliminary step to simultaneously segment and localize 5 brain structures (lateral ventricles, CSP, thalamus, cerebellum, and cisterna magna). The output of the 3D U-Net is projected in three standard planes (axial, sagittal and coronal), which are then processed by three CNNs, one for each standard plane. Each CNN outputs the 2D segmentation mask of the 5 structures. The 3D bounding boxes for the structures are reconstructed from the 2D predictions through backward projection. A total of 285 3D US volumes is partitioned following five-fold cross-validation, achieving an $IoU$ of 0.63.
%
A 3D U-Net is also used in \cite{Wyburd_2020cortical}, \cite{Venturini_2020}, \cite{hesse2020improving} and \cite{yang2020hybrid} . In~\cite{Wyburd_2020cortical}, it is used to segment the fetal cortical plate and measure the depth of the Sylvian fissure on a dataset annotated by expert clinicians using atlas. A total of 36 volumes is used to test the network performance, which achieves a $DSC$ of 0.82. 
Similarly, in~\cite{Venturini_2020} a multi-task CNN based on 3D U-Net is used for the automatic segmentation of the white matter, thalamus, brainstem, and cerebellum. The network, trained with a dataset labeled by expert clinicians using atlas, and tested on 48 volumes achieves a segmentation performance of:  $DSC$ = 0.81, $ED$ = 2.17 mm, $HD$ = 3.80 mm for thalamus, $DSC$ = 0.82, $ED$ = 2.09 mm, $HD$ = 4.14 mm for brainstem, $DSC$ = 0.77, $ED$ = 2.42 mm, $HD$ = 4.20 mm for cerebellum and $DSC$ = 0.92, $ED$ = 2.27 mm, $HD$ = 5.93 mm for white matter.
In \cite{hesse2020improving}, the authors propose a 3D U-Net combined with active contours for CSP segmentation. Results obtained on a test set of 15 volumes are evaluated with two custom metrics. 
In~\cite{yang2020hybrid}, a 3D U-Net is trained to segment the whole fetal head. The 3D U-Net is combined with a hybrid attention scheme to enhance the feature maps. Results obtained on a test set of 50 volumes are $DSC$ = 0.96, $IoU$ = 0.92 and $HD$ = 0.46 mm.
%
%

Brain-structure segmentation is further explored and included in a multi task context in \cite{namburete2018fully}. The authors propose an approach to align 3D US brain volumes,  recovered from a stack of axial slices, to a coordinate system based on skull boundaries, eye socket location, and head pose.
A multi-task FCN is trained to address the problem of 3D fetal brain localization, structural segmentation, and alignment. Results obtained testing the model on 2D axial slices sampled 
from 140 volumes are: $Acc$ = 0.99 for superior-to-inferior brain orientation, $ED$ = 6.9 mm for eye localization performance, $IoU$ = 0.82 for brain segmentation and $HD$ = 9.3 mm for target alignment error.
A method for brain localization is also proposed in \cite{moser2020automated}, which explores the use of an end-to-end 3D CNN for automated brain localization and extraction from 3D fetal US. Differently from \cite{namburete2018fully}, which predicts the position of the brain from 2D slices extracted from 3D volumes, this is a fully 3D approach and relies on a modified 3D U-Net for brain extraction. Results obtained on a test set of 300 volumes are $ED$, $HD$, $DSC$ of 1.36 mm, 9.05 mm and 0.94, respectively.  

Other than structure segmentation and localization, researchers are also working on gestational age (GA) estimation and brain development analysis.
An approach to GA estimation from 2D trans-thalamic US images is proposed in~\cite{lee2020calibrated}. The authors propose a Bayesian Neural Network (BNN) with a VGG-16 backbone and an auxiliary regression model, which are trained to predict calibrated aleatoric and epistemic uncertainties on GA. Results obtained testing the model over 2 different datasets are: (i) $MAE$ = 12.5 days on 849 frames from the same dataset used to train the model; (ii) $MAE$ = 19.6 days on 20 frames from a different dataset. 
Methods to evaluate brain development are investigated in~\cite{Namburete2017Robust} and \cite{wyburd2021assessment}. The work in \cite{Namburete2017Robust} presents a model to predict brain maturation from 3D US scans through the use of a 3D Convolutional Regression Network. The model, evaluated in five-fold cross-validation with 121 volumes for testing, obtained an error in predicting brain maturation of 6.9 days. In \cite{wyburd2021assessment}, the authors propose a method to estimate the development of 3 brain fissures (i.e., Sylvian, Parieto-occipital and Calcarine) from 3D US volumes, by predicting the fetal GA based on their respective morphology. The regions relative to each fissure are extracted from the US volumes and passed to three separate ResNet to predict the GA of each region. The model tested on 122 volumes obtained the following results: $MAE$ = 4.1, 5.1, 4.9 days 
for Sylvian, Parieto-occipital and Calcarine fissure respectively.

As for computer-assisted diagnosis of brain pathology, in~\cite{Xie_2020} the authors propose a DL-based pipeline to support the diagnosis of brain lesion from 2D images. The pipeline relies on 3 architectures for (i) cranio-cerebral segmentation, using U-Net with dilated convolutional, (ii) image classification in normal/ abnormal, using VGG-19 pre-trained on ImageNet, and (iii) lesion localization, using Grad-CAM. The dataset includes both 2D US images and 3D volume data cropped in the axial view, for a total of 15372 normal and 14047 abnormal images. For the segmentation step, $mAP$, $Rec$ and $DSC$ of 0.98, 0.91 0.94 are obtained, respectively. The overall classification $Acc$, $Rec$, $Spec$ and AUC are 0.96, 0.97, 0.96 and 0.99, respectively. The lesion localization, in the form of Grad-CAM activation heatmaps, is visually evaluated by a doctor, resulting in heatmaps localized correctly in the 61.60\% of the images. 




\subsection{Placenta \& amniotic fluid} \label{sec:placenta}

The placenta is an organ that provides oxygen and nutrients to the fetus, ensures thermo-regulation, and removes waste products from the fetus’s blood. This structure is closely related to fetus health: abnormal placental function may affect the development of the fetus, and in severe cases, even endanger the life of the fetus. Given its relevance, several clinical assessments are performed to evaluate the placenta condition along with the fetus health. 
In this scenario, placenta segmentation may provide automatic quantification of placenta volume and morphology. 

A number of approaches performs placenta segmentation from 2D US images.
The approach proposed in \cite{hu2019automated} uses a U-Net inspired CNN. The CNN is modified adding a layer to detect acoustic shadow and improve the segmentation accuracy. The model, when tested on 205 images, obtains mean $DSC$ and $Acc$ of 0.92 and 0.93, respectively. 
Similarly, in~\cite{hu2021automatic} a U-Net inspired CNN is used for placenta segmentation. An EfficientNet is then used for the classification of normal or abnormal placenta. Classification results achieved with five-fold cross-validation over 13384 frames are: $Acc$ = 0.81, $Rec$ = 0.88, and $Spec$ = 0.65. 
A U-Net inspired architecture for placenta segmentation is also used in \cite{zimmer2020multi}. The authors introduce an auxiliary classification task, incorporating the prediction of the placental position (anterior or posterior) in the U-Net architecture and improve the segmentation accuracy. 
Results obtained training the model with images of both anterior and posterior placenta and testing on 5149 2D slices extracted from 172 volumes are: $DSC$ = 0.86 and $HD$ = 28.66 mm for anterior placenta; and $DSC$ = 0.78 and $HD$ = 35.02 mm for posterior placenta.
In \cite{oguz2018combining}, three 2D cross-sectional images manually extracted from 3D US, are used to train an encoder-decoder CNN architecture. An atlas-based joint label fusion algorithm is then applied to the CNN output to combine the three prediction and enhance segmentation performace. 
The predictions from CNN and joint label fusion are combined via a random forest model to obtain the final segmentation. A mean $DSC$ = 0.86 is obtained performing four-fold cross-validation. 

Other researchers are working on placenta segmentation from 3D US data.
The work in \cite{looney2018fully}  uses a 3D U-Net. The model is evaluated through a two-fold cross-validation on a test set of 1196 US volumes, obtaining $DSC$ and $HD$ of 0.84 and 14.6 mm, respectively.  
For the same goal, in \cite{looney2017automatic} the authors use an open-source CNN trained with labels obtained through a semi-automatic Random Walker strategy. The proposed method tested on 20 US volumes obtained a $DSC$ and $HD$  of 0.73 and 27 mm, respectively. 
The work in \cite{zimmer2019towards} proposes a three stage pipeline for whole placenta segmentation and volume estimation from multi-view 3D US volumes. The pipeline includes: (i) multi-view acquisition through a multiplexer device that fixes 3 probes in an angle of 30 degree to each other and switches the view from one probe to the other almost in real time; (ii) voxel-wise fusion to combine the multiple views; (iii) segmentation based on a 3D U-Net. The model, when tested on 12 US volumes, reaches mean $DSC$ of 0.80. 
The work in \cite{yang2017towards} proposes an approach to automatically segment placenta, fetus and gestational sac. The first step includes a customized 3D CNN with long skip connections to segment the 3 structures. A RNN is used to add contextual knowledge and refine the semantic segmentation. Finally, a hierarchical deep supervision mechanism is used to boost the flow of information in the RNN and improve the segmentation performance. When tested on 44 volumes, the model achieves a $DSC$ and $HD$ of 0.64 and 24.54 mm, respectively, for placenta segmentation. 
\\

Amniotic fluid plays an important role in fetal well-being and development. Amniotic fluid has a myriad of functions: it protects  fetus and  umbilical cord, prevents infections and provides the necessary growth factors to allow normal development and growth of fetal organs. In the clinical practice, to assess the sufficiency of amniotic fluid quantity, the amniotic fluid index is used. This index is calculated dividing the maternal abdomen into 4 quadrants using the midline and the umbilicus, then the deepest pocket of amniotic fluid is evaluated in each quadrant. The index is given by the sum of the 4 measurements. With a view to automatize the index computation, amniotic fluid segmentation is a crucial task.  

In \cite{li2017automatic}, an encoder-decoder architecture with VGG16 as backbone is used to segment  amniotic fluid and fetal body in 2D US images. For amniotic fluid segmentation, the model tested on 400 images obtains $Acc$ and $IoU$ of 0.78 and 0.54 respectively. 
In \cite{cho2021automated}, a 2-step framework is used to segment the amniotic fluid pocket and measure the amniotic fluid index. 
The segmentation step is performed with a modified U-Net architecture, called AF-net, which combines atrous convolution and multi-scale side-input and side-output layers. 
Results obtained on a test set of 125 images are $DSC$, $Prec$, $Rec$ and $Spec$ of 0.87, 0.89, 0.87 and 0.99, respectively. 
This approach is further improved in \cite{sun2021complementary}, which proposes a dual-path network to segment the amniotic fluid volume from 2D US images. The primary path consists of AF-net, while the secondary path is an auxiliary CNN used to remove reverberation artifacts and complement the primary path prediction. The final segmentation output, obtained combining the primary and secondary path results, obtained a mean $DSC$, $Rec$, $Prec$ and $HD$ = 0.86, 0.81, 0.93 and 15.49 mm respectively, on a test set of 180 images.

 

\subsection{Others} \label{sec:other_organs}

\subsubsection{\textbf{Lungs}}

Immaturity of fetal lung development is the primarily cause of neonatal respiratory morbidity. 
Quantitative US imaging is often used as a non-invasive tool for fetal lung maturity assessment, through US visualization at 4CH level. The normal fetal lung has of very similar echogenicity to the adjacent liver, but with slightly different texture; and in presence of abnormalities either increasing or decreasing in echogenicity can be seen.
Recently, DL approaches have been introduced to tackle the challenges of lung analysis from fetal US image, especially for estimating GA.

%

The framework in \cite{Xia_2021} is based on DenseNet, which is trained performing a ten-fold cross-validation on a dataset of US images of 4CH plane acquired from 1023 pregnancies and divided in 3 classes based on GAs: class I (from 20 to 29+6 weeks), class II (from 30 to 36+6 weeks) and class III (from 37 to 41+6 weeks). 
The overall $Acc$ in classifying GA is of 0.83. $Rec$, $Spec$ are computed for each GA class resulting as: 0.91 and 0.76 for class I, 0.69 and 0.90 for class II, and 0.86 and 0.83 for class III, respectively. The AUC of each class is 0.98, 0.90 and 0.96, respectively.
%
%
For the same aim, the work in \cite{chen2020preliminary} proposes a two-stage transfer learning approach. 
A U-Net-like architecture with residual connection is first trained to recognize samples of fetal lung regions from other regions taken from the US images; then, the pre-trained U-Net is tuned on fetal lung region samples only to regress the corresponding GA, and thus its maturation degree. 
This study involves 332 patients, each with one 4CH US scan available, of which 126 US scans are used for testing, manually selecting the lung region, of which the gestational week is predicted. 
The authors achieve a $MAE$ of 1.56 weeks. 
However, this work provides only an indirect prediction of the lung maturation degree, as it relies on the direct correlation between the lung maturity and the GA, which does not always exist. 
%


\subsubsection{\textbf{Kidney}}
In-utero assessment of kidney is crucial to perform early diagnosis of renal pathologies. Poor kidney development is known to be associated with increased risk of kidney disease into adulthood. At the same time, abnormal fetal growth is associated to a reduced functionality of kidneys after birth \cite{Luyckx_2015, senra2020kidney}.
%
%
In this field, the work in \cite{Weerasinghe_2020} proposes a 3D U-Net to perform kidney segmentation from 3D B-mode and Power Doppler volumes. 
When tested on 20 3D images, the algorithm achieves average $DSC$, $IoU$ and $HD$ of 0.81, 0.69 and 8.96 mm, respectively.
Despite the small number of testing volumes, this is an interesting study that may pave the way for the development of early diagnosis tools for fetal kidney.

\subsubsection{\textbf{Spine}}
Fetal spine length provides insights into the fetal growth as it is affected by a variety of malformations (spina bifida, meningocele, diastematomyelia, vertebral segmentation anomalies, sacral agenesis, spinal dysgenesis, spondylothoracic or spondylocostal dysplasia).
To provide an assessment of fetal spine, the work in \cite{franz2021deep} focuses on the identification of spine centerline from 3D US scans. 
Spine segmentation is performed with a CNN that processes images at multiple scales and different fields of view. The prediction is used as input for a model-based tracing algorithm responsible to draw spine centerline. 
The model is trained in a five-fold cross-validation leaving out 80 scans for testing. Results are graphically reported for a variety of dilation radii, used for constructing ground truth masks from the annotated spine centerlines, thus no mean or best values are provided. 
%
For a similar aim, the work in \cite{chen2021neural} presents an approach to identify the spina-bifida and segment it, which is based on a U-Net model modified introducing Octave features to reduce redundant information. 
Results show that $Rec$, $Prec$, $Acc$, and $ IoU$ are 0.93, 0.96, 0.94, and 0.91, respectively; in addition, the mean standard error of the model was 4.12 mm, and its average running time reaches 12.15 seconds. This approach achieves high recognition accuracy, good segmentation accuracy, and short running time, but it is limited on few cases, without dividing among different spina bifida types and with an imbalanced presence on malformation cases compared to healthy ones (24 cases of spina bifida over a total of 3300 cases).

\subsubsection{\textbf{Abdomen}}

Fetal abdomen assessment is performed to evaluate important prognostic parameters of neonatal morbidity and mortality and to assess fetal growth. Segmentation of the abdominal outline is of interest for measuring fetal abdominal circumference. For this purpose, the work in \cite{schmidt2017abdomen} develops an approach based on CNNs to extract image features to be integrated in a deformable model. The method tested on 42 3D US images achieves a $MAE$ of 2.24 mm. This approach represents a first attempt to use DL for abdominal segmentation form 3D fetal US. However, it still relies on deformable models.
Fetal abdomen localization is achieved in \cite{droste2019towards} using an encoder-decoder architecture for US video saliency prediction. The encoder part consists of a truncated SonoNet, while the spatio-temporal decoder of the network is made up of a bidirectional gated-recurrent-unit recurrent convolutional network (GRU-RCN). In a five-fold cross-validation with 3 test videos, the architecture achieves AUC by Judd and $D_{KL}$ of 0.87 and 2.16, respectively.

\begin{figure*}[tbp]
    \centering
    \includegraphics[width = .9\textwidth]{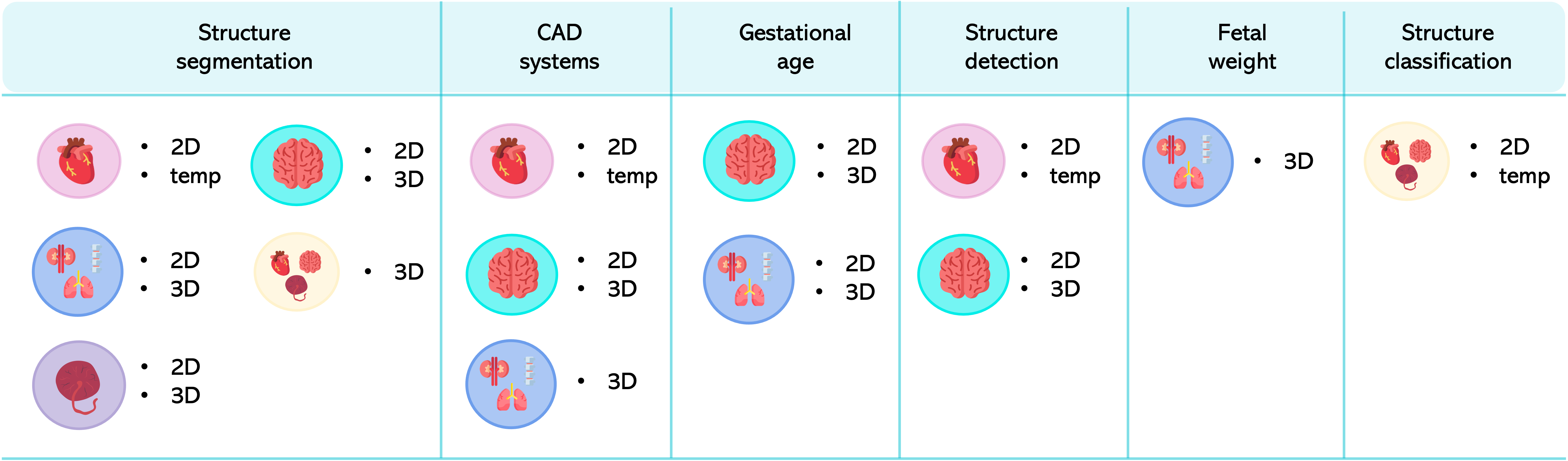}
    \caption{ Most addressed tasks (from left to right) in the field of anatomical-structure analysis. temp = temporal information }
    \label{fig:cluster_organ}
\end{figure*}

\subsubsection{\textbf{Femur}}

Volume and length of fetal femur have unique importance in fetal weight estimation. However, it is hard to be evaluated through US, due to the difficulty in locating the tips of the femur, boundary deficiency and ambiguity for tissues’ low contrasts, and variation of pose, shape and size of this structure. 
In \cite{wang2019joint}, the authors develop a unified framework for simultaneous segmentation and landmark localization of fetal femur in prenatal US volumes: fetal femur ROI is first identified through a U-Net model, then segmentation and landmark localization branches receive the common features of ROI extracted by the shared layers and generate task-specific descriptors.
The method tested on 20 US volumes reaches $DSC$, $IoU$, $HD$ and $ED$ equal to 0.91, 0.83, 4.08 mm and 0.87 mm, respectively.

\subsubsection{\textbf{Skull and Head}}
The evaluation of fetal head is a critical part of sonographic examination. This evaluation is dependent on operator experience and US image intrinsic characteristics. To attenuate these issues, DL algorithms in the literature mainly focus on the segmentation of the fetal head.
The work in \cite{cerrolaza2018deep} proposes a two-stage approach for skull segmentation in fetal 3D US. A 3D U-Net is used to roughly segment the skull. This segmentation concatenated with two additional channels (i.e., the US-wave incidence angle map and the US shadow casting map) and fed to a second 3D U-Net. 
The model, when tested on 14 volumes, achieves $DSC$ and $IoU$ of 0.83 and 0.70 respectively.
The work in \cite{perez2020deep} aims at merging several partially occluded US volumes, acquired by placing the US transducer at different projections of the fetal head, to compound a new US volume containing the whole brain anatomy. For this aim, the authors propose a pipeline of 4 CNNs. The first 2 CNNs follow what is done in \cite{perez2020deep}, while the additional 2 CNNs perform US-volume registration.
For the segmentation tasks, the pipeline achieves $DSC$ and AUC of 0.81 and 0.82, respectively, performing four-fold cross-validation on 8 US volumes.
%
A different approach is proposed in~\cite{wang2020annotation}, with the aim of segmenting fetal head. A CycleGAN is trained with a set of unpaired images and auxiliary masks obtained from a shape prior model, to generate the pseudo labels corresponding to each of the training images. The CycleGAN is equipped with a Variational Auto-encoder based discriminator and a Discriminator-guided Generator Channel Calibration module which calibrates the pseudo label generator using the discriminator’s feedback to improve the pseudo labels. The model, when tested on 200 images from the HC18 dataset, obtains a $DSC$ of 0.94.

\subsection{Multi-organ analysis} \label{sec:multiple_organs}

DL researchers are working on multi-organ analysis to reproduce the actual clinical screening, which accounts for a comprehensive list of anatomical structures.

A number of DL learning algorithms in the field focuses on classification tasks.
The work in \cite{sridar2019decision} proposes a method to automatically classify 14 different fetal structures (abdomen, arm, blood vessels, cord insertion, face, femur and humerus, foot, genitals, head, heart, kidney, leg, spine, hand) from US images by fusing information from both cropped regions of fetal structures and whole images. Two CNNs pre-trained on ImageNet are used as feature extractors. The features are classified by means of support vector machines. When tested on 965 images, the method achieves mean $Acc$, $Prec$ and $Rec$ of 0.97, 0.76 and 0.75, respectively. 
Similarly, the work in~\cite{ishikawa2019detecting} fine-tunes a VGG16 to classify US images in four classes (head, body, leg and other). This is needed as preliminary step towards the estimation of fetal position. Based on the output of Grad-CAM, the body parts position in the images is estimated. 2000 US images are used to test the algorithm obtaining an average $Rec$ of 0.92. 
Spatio-temporal classification is performed in \cite{sharma2019spatio}, in which a LSTM is used to classify 11 anatomical categories (heart, background, brain with skull and neck, Doppler maternal anatomy, spine, abdomen, nose and lips, kidneys, face side profile, femur, other). The authors use 
31629 frames from 3 US videos to test the model, achieving an $Acc$ of 0.77. 

In \cite{alsharid2020curriculum}, the authors propose a method for multi-organ (abdomen, head, heart, spine) classification from the integration of fetal US images with corresponding textual descriptions. 
The authors exploit a curriculum learning approach to train a NLP-based fetal US image captioning model with a dataset prepared using real-world US videos along with synchronized and transcribed sonographer speech recordings: 12808 and 9979 image-caption pairs are used for training and testing sessions, respectively.
The method obtains $Prec$, $Rec$ and $DSC$ equal to 0.96, 0.96 and 0.96, respectively.

An unsupervised approach to multi-organ classification is proposed in \cite{chen2020cross}, which develops a cross-device and cross-anatomy adaptation network to classify heart, abdomen and skull of an unlabelled single-sweep video dataset guided by knowledge of a labelled free-hand scanning protocol video dataset. 
The  network consists of encoder, projection layer, anatomy classifier, domain classifier and two mutual information discriminators. 
As results, the $Acc$ is reported on the test target domain data considering a 3-class classification task (without the background) and a 4-class classification task obtaining 0.73 and 0.89 for each task, respectively. 

A different approach is used in \cite{xu2018less}, where a multi-task learning framework is proposed to classify 11 different views related to abdominal organs and detect 14 landmarks from kidney, liver and spleen. This framework relies on a shared ResNet encoder and two branches, for classification and landmark detection.
For the classification, the method obtains an $Acc$ of 0.85, while for landmark detection, the average $MAE$ is 5.6 mm.

Multi-organ analysis also involves DL algorithms for localization and segmentation tasks.
In a weakly supervised fashion, the work in \cite{gao2019learning} proposes a model consisting in a CNN for extracting features and an attention-gated LSTM to localize skull, abdomen and heart in consecutive frames, considering also non-standard planes. The training is performed on fetal US videos of healthy subjects in a five-fold cross-validation.
The localization is achieved with an average $AP$ equal to 0.91. 
A pipeline to segment fetal head and abdomen is proposed in~\cite{wu2017cascaded}. The pipeline consists of three cascaded CNNs and is evaluated by means of 236 fetal head and 505 fetal abdomen images. Mean $DSC$ and $IoU$ of 0.97 and 0.96 are obtained, respectively.
In \cite{Yang2019TowardsAS}, a framework to simultaneously segment fetus, gestational sac, and placenta is proposed. The framework consists of a 3D CNN, which explores spatial intensity concurrency, and a RNN, which encodes spatial sequential to improve boundary refinement. The test over a 44 volumes dataset results in $DSC$ and $HD$ in average equal to 0.80 and 14.13 mm, respectively.

\begin{figure}[tbp]
    \centering
    \includegraphics[width = .5\textwidth]{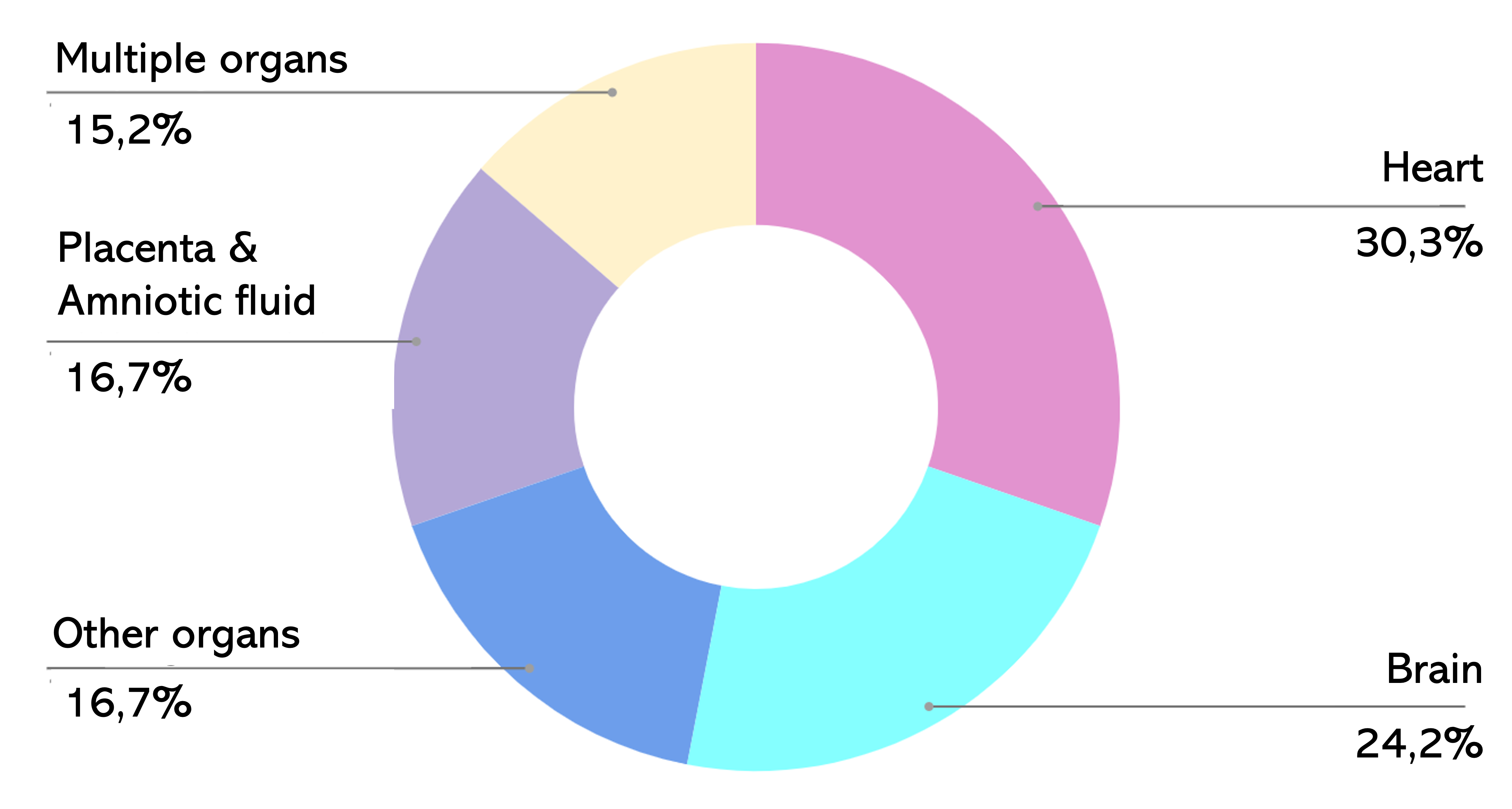}
    \caption{Number of papers surveyed in Sec. \ref{sec:organs} according to the anatomical structure.}
    \label{fig:fetal-organ}
\end{figure}

\setcounter{table}{5}

\setlength{\tabcolsep}{6pt} 
\renewcommand{\arraystretch}{1.5} 

\begin{table*}[tbp]
\scriptsize
    \centering
    \caption{Summary of deep-learning algorithms for biometry parameter estimation (for performance metrics refer to Table \ref{table:measures}). HC: Head circumference, BPD: Biparietal diameter, OFD: occipito-frontal diameter, AC: Abdominal circumference, FL: Femur length, TCD: transcerebellar diameter  }
    \label{tab:biometry}
    
    \begin{tabularx}{\textwidth}{lXXXlX}

        Paper (Year) & Biometry & Training set size & Test set size & Annotators & Performance metrics \\
        \hline
        
        Sinclair et al., \cite{sinclair2018human} (2018) & HC, BPD  & 2164 2D & 100 2D & 2 & $DSC$ = 0.98, $MAE$ = 1.3 mm \\
        
        Rong et al., \cite{rong2019deriving} (2019) & HC & 999 2D (HC18) & 335 2D (HC18) & - &  $DSC$ = 0.95, $MAE$ = 2.44 mm \\ 
        
        Skeika et al., \cite{skeika2020convolutional} (2020) & HC & 999 2D (HC18) & 297 2D (HC18) & 1 &  $DSC$ = 0.97, $MAE$ = 1.89 mm \\
        
        Zeng et al., \cite{zeng2021fetal} (2021) & HC & 999 2D (HC18) & 335 2D (HC18) & Few &  $DSC$ = 0.98, $MAE$ = 1.77 mm \\ 
        
        Aji et al., \cite{aji2019automatic} (2019) & HC & 999 2D (HC18) & 335 2D (HC18) & Few & $DSC$ = 0.76, average error = 14.96\%\\

        Qiao et al., \cite{qiao2020dilated} (2020) & HC & 849 2D (HC18) & 150 2D (HC18) & - & $DSC$ = 0.97, $MAE$ = 2.27 mm \\ 
        
        Oghli et al., \cite{oghli2020automatic} (2020) & HC & HC18 & HC18 & 1 & $DSC$ = 0.95, $HD$ = 4.5 mm \\ 
        
        Bhalla et al., \cite{bhalla2021automatic} (2021) & HC & HC18 & HC18 & - & $MAE$ = 2.16 mm \\ 
        
        Sobhaninia et al., \cite{sobhaninia2019fetal} (2019) & HC & HC18 & HC18 & - & $DSC$ = 0.97, $MAE$ = 2.12 mm \\ 
        
        Sobhaninia et al., \cite{sobhaninia2020localization} (2019) & HC & 849 2D (HC18) & 150 2D (HC18) & - & $DSC$=0.92, $MAE$ = 2.22 mm \\
        
        Li et al., \cite{li2020automated} (2020) & HC, BPD, OFD & 999 2D (HC18) & 335 2D (HC18) & Few & $DSC$ = 0.97, $MAE$ = 1.81 mm \\ 

        Kim et al., \cite{kim2019automatic} (2019) & HC, BPD & 102 2D & 70 2D & 2 & $ACC$=0.87, success rate = 0.93\\ 
        
        Al et al., \cite{al2019improving} (2019) & HC & 999 2D (HC18) & 335 2D (HC18) & 1 & $DSC$ = 0.97, $MAE$ = 2.33 mm \\ 
        
        Fiorentino et al., \cite{fiorentino2021regression} (2021) & HC & 999 2D (HC18) & 335 2D (HC18) & 1 & $DSC$ = 0.97, $MAE$ = 1.90 mm \\
        
        Moccia et al., \cite{moccia2021mask} (2021) & HC & 999 2D (HC18) & 335 2D (HC18) & 1 & $DSC$ = 0.98, $MAE$ = 1.95 mm \\
        
        Meng et al., \cite{meng2020cnn} (2020) & HC & 905 2D (HC18) & 94 2D (HC18) & - & $DSC$ = 0.97 \\
        
        Budd et al., \cite{budd2019confident} (2019) & HC & 2848 2D (2000 subjects*) & 540 2D (2000 subjects*) & 45 & $DSC$ = 0.98, $MAE$ = 1.81 mm \\
        
       Zhang et al., \cite{zhang2020direct} (2020) & HC & 800 2D (HC18) & 199 2D (HC18) & - & $MAE$=4.52 mm\\ 
       
       Zhang et al.,\cite{zhang2020explainability} (2020) & HC & 800 2D (HC18) & 199 2D (HC18) & - & $MAE$ = 4.78 mm \\ 
       
       Jang et al., \cite{jang2017automatic} (2017) & AC & 56 2D (1 subject*) & 32 2D (1 subject*) & 2 & $DSC$=0.85, $ACC$=0.79 \\ 
       
       Kim et al., \cite{kim2018machine} (2018) & AC & - (77 subjects*) & - (77 subjects*) & - & $DSC$= 0.92, $ACC$=0.87\\  %
       
       Cengiz et al. \cite{cengiz2021automatic} (2021) & CRL & 697 2D & 545 2D & 1 & $DSC$ = 0.93, $IoU$ = 0.88, $MAE$ = 3.87 mm\\
       
       Chen et al., \cite{Chen_2020LV} (2020) & LV & 4379 2D & 500 2D & 3 & $Acc$=0.96, $Prec$=0.98, $Spec$=0.90, $MAE$= 1.80 mm \\
       
       Zhu et al., \cite{zhu2021automatic} (2021) & FL & 2300 2D (435 subjects*) & 310 2D (435 subjects*) & 1 & $DSC$=0.92, $MAE$= 0.46 mm \\ 
       
       Chen et al., \cite{chen2021ellipsenet} (2021) & CTR & 1669 2D & 417 2D & 2 & $DSC$ = 0.93\\
        
       Bano et al., \cite{bano2021autofb} (2021) & HC, AC, FL & 262 2D (42 subjects*) & 87 2D (42 subjects*) & 1 & $MAE$ = 2.67 mm (HC), $MAE$ = 3.77 mm (AC), $MAE$ = 2.10 mm (FL) \\ 
        
       Plotka et al.\cite{plotka2021fetalnet} (2021) & HC, AC, FL & 274275 2D (560 patients) & 57001 2D  (140 patients) & 6 & $DSC$ = 0.96, $MAE$ = 2.9 mm (HC), $MAE$ = 3.8 mm (AC), $MAE$ = 0.8 mm (FL) \\ 
        
       Gao et al., \cite{gao2021dual} (2021) & HC, TCD & 937 2D (937 subjects) & 913 2D (913 subjects) & 1 & $MAE$=2.04 mm\\ 

    \end{tabularx}

\end{table*}

\setcounter{table}{5}
\begin{table*}[tbp]
\scriptsize
    \centering
    \label{tab:biometry}
    
    \begin{tabularx}{\textwidth}{lXXXlX}

        Paper (Year) & Biometry & Training set size & Test set size   & Annotators & Performance metrics \\
        \hline
        
        Ryou et al., \cite{ryou2019automated} (2019) & CRL, HC, AC & 44 3D & 21 3D & Few & $IoU$ = 0.92 for abdomen and head and a $IoU$ = 0.66 for limbs segmentation, $MAE$ = 2.24 mm\\ 
       
       Oghli et al., \cite{oghli2021automatic} (2021) & AC, BPD, FL, HC & 999 2D (HC18) + 1154 2D (551 subjects*) & 335 2D (HC18) + 180 2D (551 subjects*) & 2 & $DSC$ = 0.98, $HD$ = 1.14 mm, average perpendicular distance = 0.2 mm \\ 
       
       Prieto et al., \cite{ prieto2021automated} (2021) & BPD, HC, CRL, AC, FL & 147855 2D (4433 subjects) & 7233 2D (2491 subjects) & Few & $Acc$ = 0.93, $IoU$ = 0.91, $MAE$=1.89 cm \\
       
       Rasheed et al., \cite{rasheed2021automated} (2021) & HC, BPD & 30000 2D & 1000 videos & 1 & $Acc$=0.96.
       
    \end{tabularx}

\end{table*}



%

\subsection{Limitations and open issues for anatomical-structure analysis}

As shown in Fig. \ref{fig:cluster_organ}, segmentation is the most addressed task for all the anatomical structures. However, most of the proposed methodologies consist of 2D U-Net-based architectures, while more advanced architectures exploited in closer fields (e.g. adversarial segmentation, spatio-temporal processing) are investigated less. 
%
Interestingly, priors relevant to anatomical-structure shape have not been fully exploited yet. However, shape-constraint DL strategies have being recently proposed in the literature to drive the output of segmentation CNN towards the desired structure shape, for example by using adversarial learning~\cite{bohlender2021survey,casella2021shape}.

However, newer techniques that have become more widely available, such as volume sonography (3D/4D/spatiotemporal image correlation (STIC)) have not been exploited yet for a more detailed anatomical and functional assessment of the fetal heart.

Up to now, the most studied anatomical-structures are heart and  brain, contributing with the 30.3\% and 24.2\% of the surveyed papers, respectively, as shown in Fig. \ref{fig:fetal-organ}. Few papers focus the attention on multi-organ analysis: the inter-organ relations are frequently exploited by doctors when navigating and interpreting medical images \cite{cerrolaza2019computational}, making the multi-organ analysis essential for a good fetus evaluation. The lack of annotated dataset even in this field, does not allow to 1) characterize the complex inter-organ relations and 2) underline the difference between pathological and physiological images since data which mix health and pathological cases is critical to develop CAD systems robust to pathology and unusual anatomy.


The number of applications for DL in fetal brain anatomical-structure analysis are heterogeneous and each one focuses on a particular task, such as brain localization, segmentation, classification, and GA estimation. Given the fragmented nature of the literature available in this field, comparing the different approach is challenging. The lack of public datasets further hampers a fair comparison among the different approaches.


\begin{figure}[tbp]
\centering
\subfigure[\label{subfig:upconv1}] 
   {\includegraphics[width=.23\textwidth]{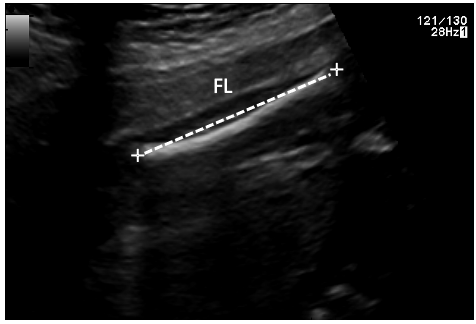}}
\subfigure[\label{subfig:upconv2}]
   {\includegraphics[width=.23\textwidth]{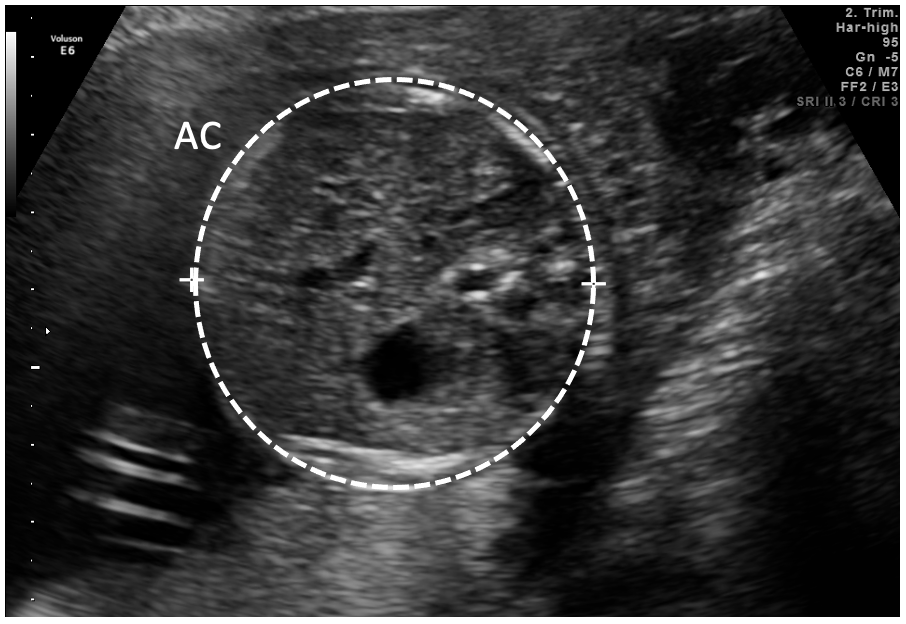}}
\subfigure[\label{subfig:upconv3} ]
   {\includegraphics[width=.23\textwidth]{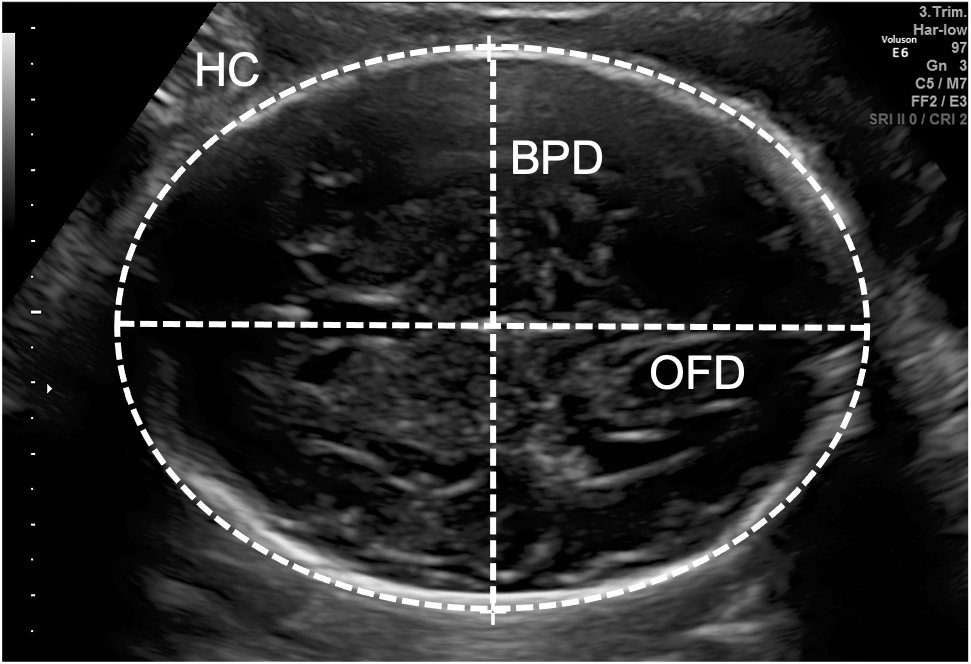}}
  \subfigure[\label{subfig:upconv4} ]
   {\includegraphics[width=.23\textwidth]{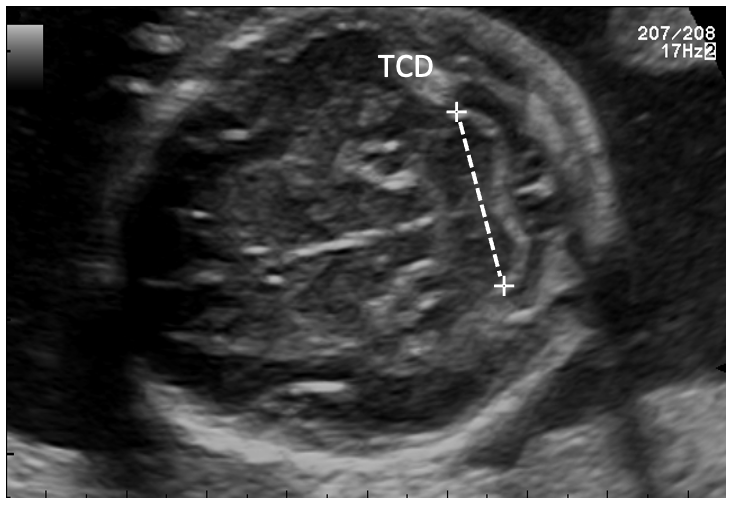}}
 \caption{\label{fig:biometry} 
 Common biometry parameters.
Transverse section of: (a) femur length (FL), (b) abdominal circumference (AC), (c) head circumference (HC), biparietal diameter (BPD) and occipito-frontal diameter (OFD), (d) transcerebellar diameter (TCD). }
\end{figure}

\section{Biometry parameter estimation}
\label{sec:biometry}

Assessing fetal size and GA, as well as detecting fetal growth abnormalities, lay the foundation of modern prenatal care. 
Fetal biometry assessment represents the most common medical investigation undertaken in this regard. 
Before 14 weeks, the GA and fetal size are estimated by the measurement of crown-rump length (CRL), which is calculated from the top of fetus's head to bottom of its torso. Once the CRL exceeds a fixed length (generally after 14 weeks), common measurements include HC,  biparietal diameter (BPD),  occipito-frontal diameter (OFD),  transcerebellar diameter (TCD), lateral ventricles (LV),  abdominal circumference (AC) and  femur diaphysis length (FL). 
Figure~\ref{fig:biometry} shows image samples for each of these biometries.
These fetal biometries are used to assess fetal growth trajectory and ensure normal fetal development when measured at different points in time (trimesters).
Additionally, cardio-thoracic ratio (CTR) and cardiac axis biometrics are measured for diagnosing CHD.

Biometry measurements are performed in a standardized manner by identifying the appropriate sonographic plane and by precisely placing calipers in the corrected position \cite{salomon2019isuog}. 
Over the years, DL algorithms that provide automated placement of the calipers to measure BPD, OFD, CRL, TCD, HC and AC have been extensively exploited to reduce operator-dependent errors and improve accuracy of fetus well-being assessment.   Table \ref{tab:biometry} summarizes the papers surveyed in this section.

Most of the approaches in literature tackle the problem of caliper measurement by segmenting the area of interest at first. 
In \cite{sinclair2018human}, a VGG-like CNN is used to segment fetal head and successively provide HC and BPD measurements by means of ellipse fitting. A total of 100 test images annotated by two experts is used to evaluate the proposed method. Mean $DSC$ and $MAE$ of 0.98 and 1.3 mm are obtained, respectively.

More recently, encoder-decoder  CNNs  have been investigated for the segmentation task. 
Here, the majority of published papers focuses on HC measurement only, relying on the HC18 challenge dataset (Sec. \ref{sec:dataset}). 
The work in \cite{rong2019deriving} proposes an active-contour model guided by external forces that are derived with a U-Net-like CNN trained to segment the fetal head. The HC is then identified via ellipse fitting with the direct least squares fitting \cite{fitzgibbon1999direct}. Mean $DSC$ and $MAE$ of 0.95 and 2.44 mm, respectively, are obtained. 
For the same goal, in \cite{skeika2020convolutional} a V-Net is used to automatically segment the fetal skull and the HC is extrapolated, obtaining $DSC$ and $MAE$ 0.97 and 1.89 mm, respectively. 
V-Net is modified to encode attention mechanism in~\cite{zeng2021fetal}, obtaining $DSC$ and $MAE$ of 0.98 and 1.77 mm, respectively. 
A U-Net architecture is instead used in \cite{aji2019automatic}, reaching a $DSC$ = 0.76 and an average error = 14.96\%.
Similarly, in~\cite{qiao2020dilated} a U-Net architecture is modified by adding dilated convolution layers after the last layer of the encoder and squeeze-and-excitation blocks on the skip connections. The method reaches a $DSC$ and $MAE$ of 0.97 and 2.27 mm, respectively. 
In \cite{oghli2020automatic} a multi-feature pyramid U-Net is proposed obtaining $DSC$ and $HD$ of 0.95 and 4.5 mm, respectively. 
An encoder-decoder network is also proposed in~\cite{bhalla2021automatic}. The network consists of dense blocks that strengthen feature propagation and channel attention in bottleneck to enhance the important features. A $MAE$ = 2.16 mm is obtained. 
In all the previous work HC is extrapolated by means of ellipse fitting methods.

The work in \cite{sobhaninia2019fetal} proposes a multi-task CNN inspired by the structure of LinkNet, for the automatic segmentation of fetal head and the successively estimation of the HC main axes, center and angle. The approach is evaluated on the HC18 challenge test, obtaining mean $DSC$ and $MAE$ of 0.97 and 2.12 mm, respectively. The authors present another approach~\cite{sobhaninia2020localization} based on a multi-scale and low complexity structure inspired by LinkNet. Mean $DSC$ and $MAE$ of 0.92 and 2.22 mm, respectively, are obtained. 
A multi-task approach is also exploited in \cite{li2020automated}, where head segmentation accuracy is enhanced by a feature pyramid inside a U-Net-like CNN. A regressor branch is added for predicting the HC value. $DSC$ and $MAE$ of 0.97 and 1.81 mm are obtained. 

To improve segmentation accuracy, some papers incorporate a localization network in the pipeline to relieve the segmentation network from learning the position of the head.
In~\cite{kim2019automatic}, a region proposal network is used as post processing procedure once the fetal head is segmented by U-Net on polar transformed images. HC and BPD are extrapolated by an ellipse fitting method. Even if with encouraging results ($ACC$=0.87 for plane acceptance check and success rate = 0.93 for HC and BPD estimation), the approach is tested on 70 images only on a custom dataset. 
A Mask-RCNN is used on \cite{al2019improving} to improve segmentation accuracy by incorporating a object localisation framework. Coordinates are extrapolated by means of a ellipse fitting method. The proposed model is tested on the HC18 dataset, achieving $DSC$ and $MAE$ of 0.97 and 2.33 mm, respectively.

A different strategy is used on \cite{fiorentino2021regression}, which considers the problem of fetal head caliper placement as a distance regression task. The proposed framework consists of a tiny-YOLO for head location and centering, and a regression CNN, which by means of distance field, accurately delineates the fetal head boundaries. HC is successively obtained using a ellipse fitting method. The framework is evaluated on the HC18 challenge reaching a $DSC$ and $MAE$ of 0.97 and 1.90 mm, respectively. The work is further improved in \cite{moccia2021mask} in which the framework is made end-to-end using a Mask-RCNN along with ellipse fitting for HC extrapolation. Results are comparable ($DSC$ = 0.98, $MAE$ = 1.95 mm).
The direct regression of the contours coordinates is also performed in \cite{meng2020cnn}. However, CNN is jointly with a attention refinement module along with a graph convolution network (GCN). This approach does not requires any ellipse fitting method. A $DSC$ = 0.97 is obtained when testing on the HC18 dataset.

\begin{figure}[tbp]
\centering
\subfigure[\label{subfig:seg} Segmentation (along with possible head localization pre-prossing) + Ellipse fitting]
   {\includegraphics[width=.33\textwidth]{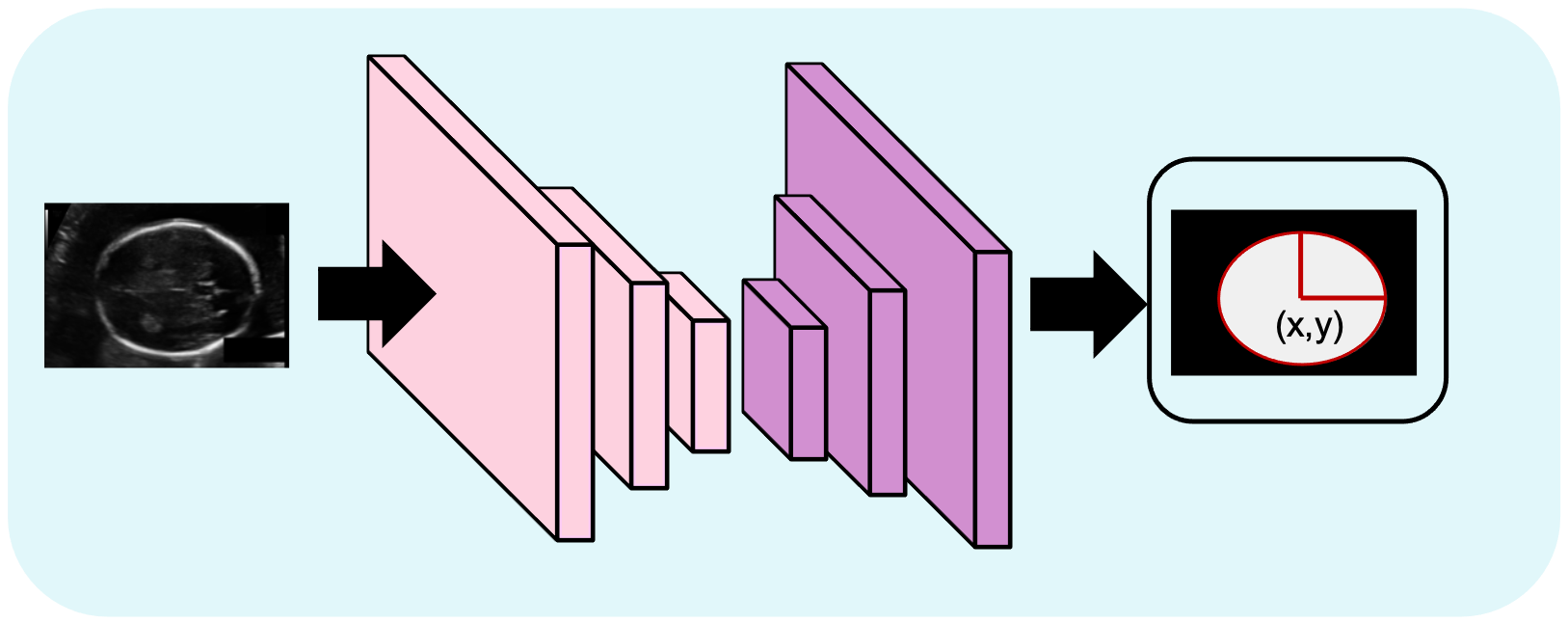}}
\subfigure[\label{subfig:reg}Region proposal along with boundary regression + Ellipse fitting]
   {\includegraphics[width=.33\textwidth]{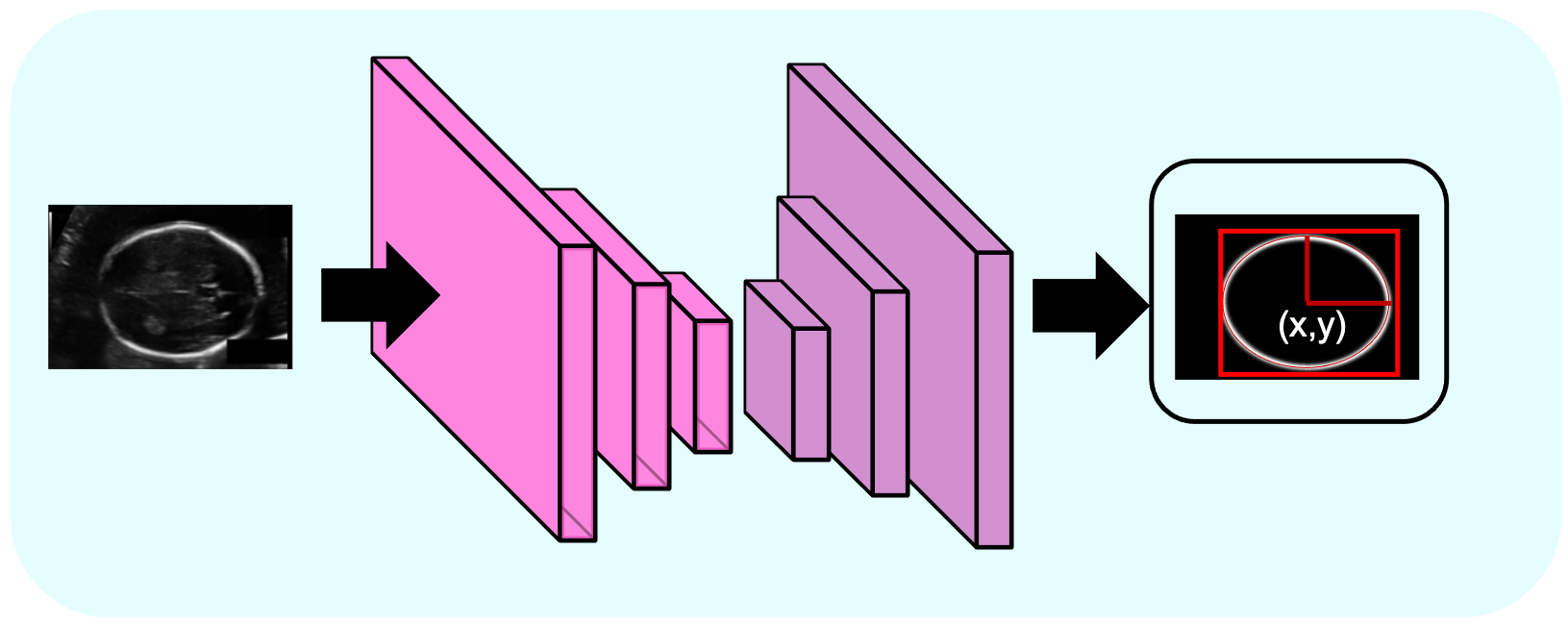}}
\subfigure[\label{subfig:multi} Multi-task framework ]
   {\includegraphics[width=.33\textwidth]{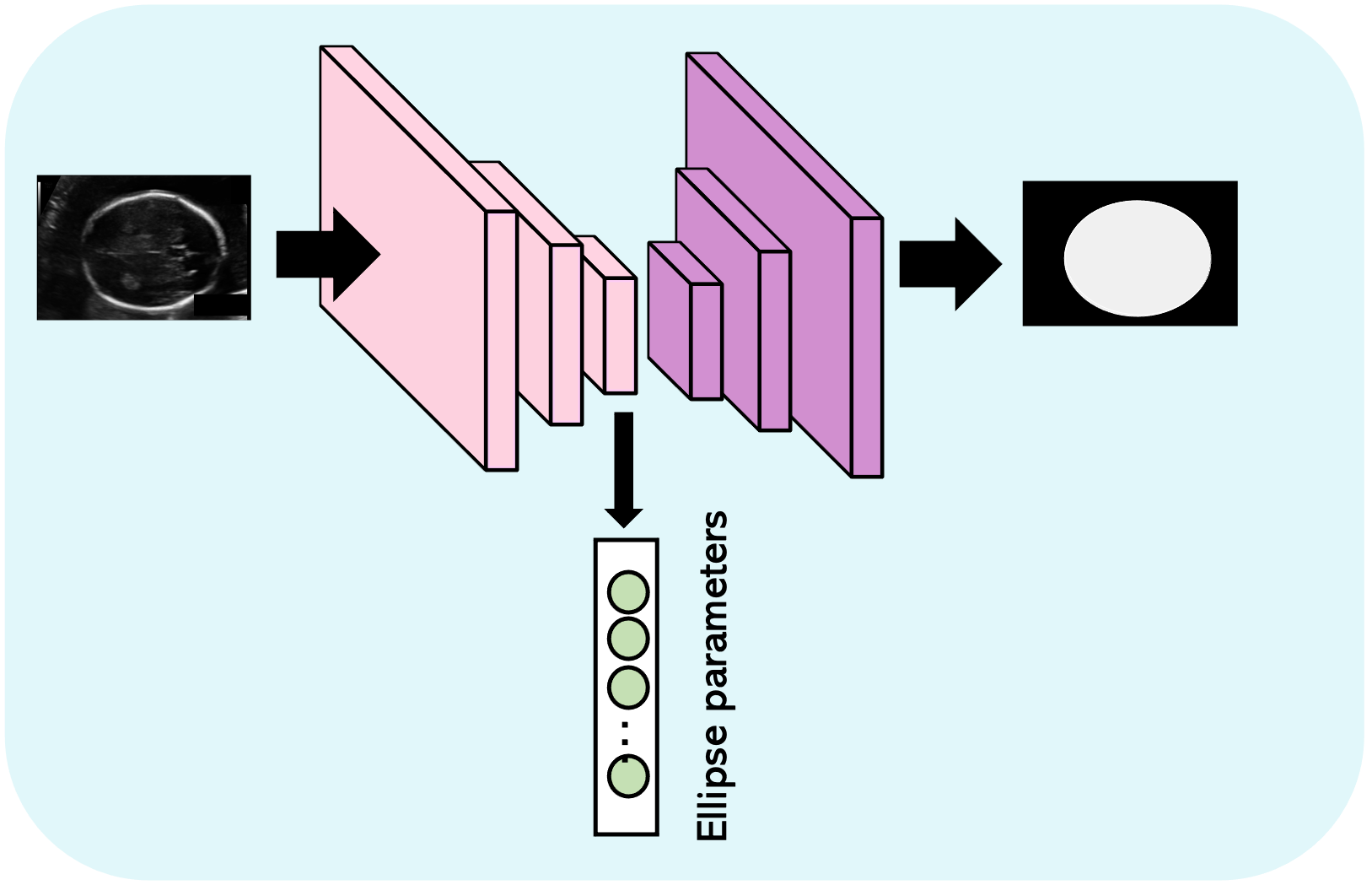}}
\subfigure[\label{subfig:regg} Direct regression ]
   {\includegraphics[width=.25\textwidth]{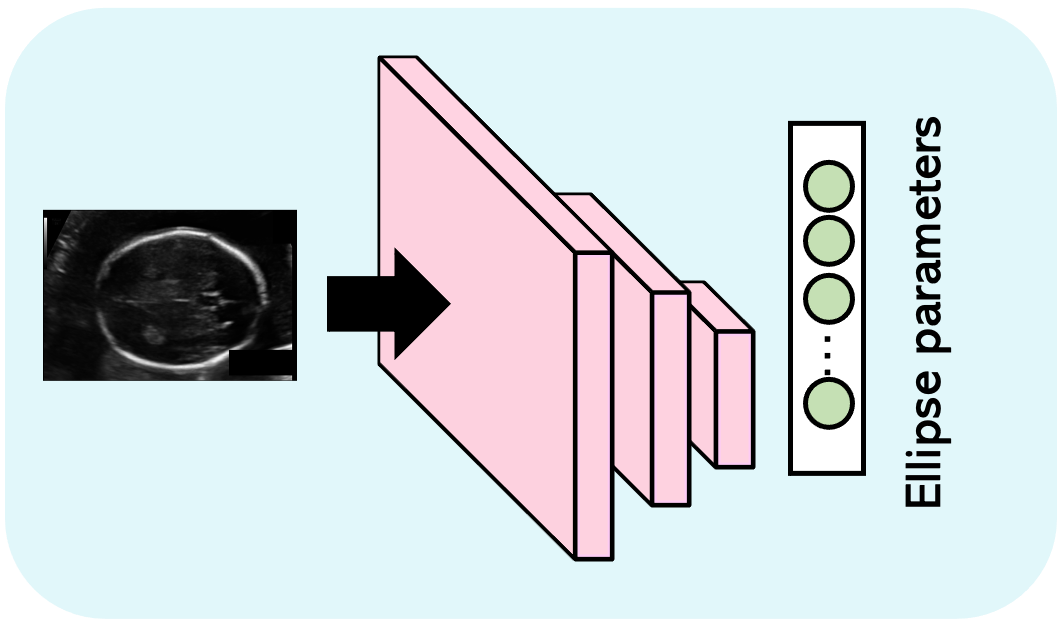}}
 \caption{\label{fig:overview_biometry} Most common deep-learning strategies to fetal HC measurement. }
\end{figure}

The work in \cite{budd2019confident} further extends the research in the field of HC estimation by developing two probabilistic CNN methods: Monte Carlo Dropout during inference and a probabilistic U-Net. These methods are particularly useful in the clinical practice since multiple plausible semantic segmentations of fetal heads along with HC measurements are provided to the clinicians, which can choose the best option. Mean $DSC$ and $MAE$ of 0.98 and 1.81 mm are obtained from a custom dataset of 540 test images labelled by 45 expert sonographers.

In order to avoid intermediate steps, such as segmentation CNN along with ellipse fitting methods, which may be computationally expensive (both for model training and labeling), the work in \cite{zhang2020direct} proposes a regression CNN to directly estimate the HC measure. Different CNNs are tested (including VGG16 and ResNet) as backbones, reaching a $MAE$ of 4.52 mm with data from the HC18 challenge. A similar approach is followed in~\cite{zhang2020explainability} in which saliency maps are used to provide interpretation of the  regression CNN results. 

Besides head biometries, a number of researchers is focusing on AC estimation. 
The work in \cite{jang2017automatic} proposes a method for the AC estimation from 2D US data acquired from 88 patients. The framework performs semantic segmentation of several anatomical structure (i.e., stomach bubble, amniotic fluid, and umbilical vein) by using a custom designed CNN. AC measurement is achieved through Hough transformation, and a plane acceptance check performed by another CNN, which verifies the presence of all structures of interest.
The $DSC$ obtained over the test set is equal to 0.85 for AC measurement and the mean $ACC$ for the plane acceptance check is 0.79. 
This method faces problems in predicting the AC precisely in the case of insufficient amniotic fluid, which commonly occurs when observing oversized fetuses.
To improve the results achieved by \cite{jang2017automatic}, the work in \cite{kim2018machine} proposes a method consisting of a combination of multiple CNNs.
A CNN is used to identify anatomical structures in US images (stomach bubble, amniotic fluid, and umbilical vein) acquired from 77 pregnant women and Hough transform is used to obtain an initial estimate of the AC. These data are fed to other CNNs to estimate the spine and bone positions, which are used to compute AC accurately. Then, a U-Net and a classification CNN are used to check whether the image is suitable for AC measurement. With this framework, a $DSC$ of 0.92 is achieved for AC measurement and an $ACC$ of 0.87 for acceptance check of the fetal abdominal standard plane. 

Only one work \cite{cengiz2021automatic} focuses on CRL measurement estimation. A U-Net is used to segment the all fetus area and CRL is computed from the obtained segmentation. A total of 545 images is used to evaluate the approach, reaching $DSC$, $IoU$ and $MAE$ of 0.93, 0.88 and 3.87 mm, respectively.

%


Few papers in the literature focus on FL, LV biometry and CTR estimation.
A first approach for automatically measuring the width of LV is proposed in \cite{Chen_2020LV}. This approach, which relies on Mask-RCNN, reaches $Acc$, $Prec$, $Spec$ and $MAE$ of 0.96, 0.98, 0.93 and 1.8 mm respectively, when tested on 500 images labeled by 3 experienced sonographers.
FL measurement is performed with SegNet and image skeletonization in~\cite{zhu2021automatic}. A total of 310 images labeled by an expert sonographer is used to validate the approach, reaching a $DSC$ and a $MAE$ of 0.92 and  0.46 mm, respectively.
The work in \cite{chen2021ellipsenet} focuses on CTR and cardiac axis estimation. The proposed CNN is a one-stage ellipse detection algorithm. The 20\% of 2086 fetal echocardiographic images is used as testing set, reaching a $DSC$ = 0.93. However, no measurements in pixels or mm are reported.

Multiple biometry estimation is addressed in \cite{bano2021autofb}, where  head, abdomen and femur are semantically segments using state-of-the-art CNNs. This step is followed by region fitting and scale recovery for the biometry estimation. The framework is evaluated through four-fold cross-validation on a dataset of 349 images (42 pregnancies). $MAE$ of 2.67 mm, 3.77 mm and 2.10 mm are obtained for HC, AC and FL, respectively. 
The work in \cite{plotka2021fetalnet} proposes a multi-task framework consisting of U-Net along with ConvLSTM to jointly localize, classify and measure HC, AC and FL in fetal US video. 
A mean $DSC$ od 0.96 is obtained, along with $MAE$ = 2.9 mm (HC), $MAE$ = 3.8 mm (AC) and $MAE$ = 0.8 mm (FL). 
A multiple biometry estimation is performed also in \cite{ryou2019automated} with the aim of supporting first trimester fetal assessment from a single 3-D US scan. The framework extracts automatically a slice of the whole fetus in the sagittal view and simultaneously segments the whole fetus by means of a multi-task network. Automated segmentation of the whole fetus into head, abdomen and limbs is then performed to extrapolate biometry measurements. A dataset consisting of 21 US volumes is used as testing set, reaching a $IoU$ = 0.92 for abdomen and head and a $IoU$ = 0.66 for limbs segmentation. A $MAE$ = 2.24 mm is obtained for head and abdomen biometry measurements.
The work in \cite{oghli2020automatic} is further expanded on \cite{oghli2021automatic} in which HC, BPD, AC and FL measurements are also considered. A total of 335 images (HC18 test set), along with 180 images coming from a local dataset, is used to test the method. $DSC$ and $HD$ of 0.98 and 1.14 mm are obtained, respectively. 

The work in~\cite{gao2021dual} proposes an approach that relies on unsupervised domain adaptation for HC and TCD biometry estimation from low cost US devices. 
The domain invariant representations for feature extraction are jointly learnt from both high-end and low cost US images and a unsupervised learning is used to calibrate the model in the low cost domain in order to produce consistent predictions. The final biometric estimation is performed fitting an ellipse to the contours (for HC measurements) and performing non-maximum suppression to find the pixels with greatest probabilities (for TCD measurements). A total of 913 images are used for model evaluation. An $MAE$ = 2.04 mm between HC and TCD measurements is obtained.

In \cite{prieto2021automated}, a fully-automated pipeline for GA estimation is proposed. The pipeline relies on BPD, HC, CRL, AC and FL estimation, which is accomplished following U-Net based structure segmentation.
For this study, three different datasets are used: the first one, including 23209 fetal US images from 1450 women, and the second one, with 124646 2D images from 2,983 US sessions, are used for training the algorithm, while the third one with 7233 US images from 2491 US studies is used for the evaluation.
An approach to automate the fetal head biometry in real-time is proposed in \cite{rasheed2021automated}. Firstly, an AlexNet is used to classify and extract fetal head from US images; successively a U-Net is  used to compute HC and BPD to estimate gestational age. A dataset consisting of 1000 US videos is used to validate the model, which achieves an $Acc$ of 0.96.


\subsection{Limitations and open issues for biometry parameter estimation}

Most of the papers dealing with biometry parameter estimation focuses the attention on HC estimation only. 
This may be attributed to the release of the HC18 dataset, which strongly advanced the research in this domain. However, the HC18 dataset does not allow to study advanced DL methodologies due to both the limited number of images and anatomical structures (only fetal brain is shown). 
%
From the virtuous example of the HC18 challenge, the research community should work to collect and release datasets for other biometries, too. 

An end-to-end approach to identify the scan plane at first, and then calculate  biometries associated to that plan, could be a valuable support tool for clinicians. Only the work in \cite{prieto2021automated} proposes something similar, but further research is needed.

A unified framework to biometry estimation from multiple anatomical regions have not been proposed, yet. Here, the most peculiar challenge is the huge variability in terms of shape and morphology of the anatomical structures (i.e. abdomen less contrasted with the background as opposed to other organs, femur and cerebellum that not fit an ellipse), as well as the dimension of the organs, which varies according to the gestational trimesters.


\section{Other tasks}
\label{sec:other}

Besides standard-plane detection (Sec. \ref{sec:plane_det}), anatomical-structure analysis (Sec.~\ref{sec:organs}) and biometry measurements (Sec.~\ref{sec:biometry}), researchers are working on a number of other tasks, including:



\subsubsection{\textbf{Adipose tissue evaluation}}
Observation and evaluation on fetal adipose tissue is significant to determine the growth and nutritional adequacy of the fetus.
The work in \cite{vaze2018segmentation} performs fetal adipose tissue segmentation through a U-Net with depth-wise separable convolutions. The performance on a test set of 340 US slices, extracted from 68 volumes, results in $Rec$, $Spec$ and $DSC$ of 0.87, 0.99 and 0.80, respectively. 

%
The work in \cite{hermawati2021phase} proposes a framework for detecting and segmenting the fetal thigh cross-sectional area of adipose tissue. The framework relies on Faster R-CNN for localizing the regions containing the fetal thigh cross-section, on which threshold-based segmentation is performed to allow measurement of adipose tissue thickness.

The framework achieves, on a test set of 50 cross-sectional US images, $Prec$, $Rec$, $Spec$ and $DSC$ of 0.92, 0.94, 0.95 and 0.93, respectively. 

\subsubsection{\textbf{Face analysis}}
Accurate detection and visualization of fetal face position and orientation is crucial in prenatal diagnosis, growth monitoring and detection of fetal anomalies. A U-Net inspired encoder-decoder 3D CNN is proposed in~\cite{singh2021deep} to segment fetal face from 3D US volumes. The model obtains a mean $ED$ of 1.72 mm over a five-fold cross-validation with 6 volumes for testing in each fold.  
In \cite{chen2020region}, an RPN-based object detection framework is proposed  to detect landmarks in 3D facial US volumes. Predictions from the RPN architecture are further refined with a distance-based graph prior to produce the final bounding box for each landmark. The model tested on 32 volumes obtained a mean $IoU$ of 0.64.

\subsubsection{\textbf{Gender identification}}

Determining fetal gender by US scan is among the main tasks during the early stages of pregnancy.
With the aim of canceling this information from the US display to prevent unauthorized gender viewing, the work in \cite{lakra2019deep} develops a residual network to identify frames containing gender-defining region in real-time among the entire set of images in cine-loop acquired from 50 women, between 11 and 20 weeks of gestation. The training is performed in five-fold cross-validation and, despite the strong dataset imbalance, in average their method achieve $Acc$, $Rec$, $Prec$ and $F1$ equal to 0.83, 0.82, 0.70 and 0.69, respectively.

\subsubsection{\textbf{Pose estimation}}
In  \cite{yang2019fetusmap}, fetal pose is estimated by localizing 16 landmarks, including joints. 
The authors propose a self-supervised learning framework to fine tune a network to form visually plausible pose predictions: a pre-trained 3D U-Net predicts the heatmaps of the 16 landmarks with an intermediate fetal pose estimator. By retrieving a support set of atlases in the pose library via rigid registration, label proxies are produced to form the self-supervision. The landmark detector is tuned iteratively for on-line refinement and updated under the gradient checkpointing strategy in necessary. 
The performance is evaluated over 52 US volumes, achieving $ED$ and AUC in landmark detection of 4.92 mm and 62.90\%, respectively.

\subsubsection{\textbf{Preterm-birth prediction}}
An early detection of risks of preterm birth is crucial to timely intervene and preserve fetal survival. US data are typically inspected by expert doctors, which use hand-designed image features such as cervical length and anterior cervical angle. Lately, to overcome errors related to a subjective and manual method, DL was used to predict preterm birth.

In \cite{wlodarczyk2019estimation}, a U-Net is used to segment the cervix from US images. The masks obtained are used to estimate the cervical length and the anterior cervical angle. The cervical length is measured through the centerline algorithm and evaluated performing linear regression between estimated and ground truth lengths of cervix. For the anterior cervical angle estimation is used an iterative algorithm that finds centroid point of the cervix mask and splits it into two shapes. In the end, the anterior cervical angle is measured between the anterior wall and the line between the last two centroids. On a test set of 108 images the segmentation step achieves an $IoU$ of 0.91; while the final classification to assign preterm vs control labels, performed with Bayes classifier in a five-fold cross-validation, obtains an $Acc$, $Prec$, $Rec$ and AUC of 0.77, 0.85, 0.74 and 0.78 respectively.
A similar approach is  proposed in \cite{wlodarczyk2020spontaneous},which uses a U-Net with a parallel branch for classification to simultaneously segment and classify the cervix in 2D US images. The model tested on 70 images reaches a $IoU$, $Rec$, $Prec$ and AUC of 0.92, 0.67, 0.68 and 0.72 respectively.

\subsubsection{\textbf{US simulation}}
US simulation based on ray tracing can provide an interactive environment for training sonographers as an educational tool. However, a trade off between image quality and interactivity has to be taken into account, as it can lead to sub-optimal results in terms of interactive rates.
The work in \cite{zhang2020deep} proposes a patch-based GAN for improving the quality of simulated US images while keeping constant computational time, via image translation of computationally low-cost images to high quality simulation outputs. In addition, segmentation and attenuation integral maps are provided to the translation framework to improve the preservation of anatomical structures and the synthesis of relevant acoustic shadows.
The dataset used consists of patches extracted from 6669 images and a constant binary mask covering the beam shape for all samples. When tested on 669 US images, the framework achieves mean KLD of 13.80.

\subsubsection{\textbf{Shadow artifact removal}}

Acoustic shadows caused by sound-opaque occluders can potentially hide vital anatomical information in 2D US and thus can be a big burden for US analysis, ranging from anatomy segmentation to landmark detection.
%
An automatic shadow detection method is presented in  \cite{meng2018automatic}. It generates a pixel-wise shadow confidence map from weakly labelled annotations, jointly using: a FCN as shadow image discriminator, a feature attribution map from a Wasserstein GAN and an intensity saliency map from a graph cut model.
The evaluation is performed on two US datasets, one containing 993 2D images from 14 fetal standard planes and the other consisting of 643 brain 2D images; the $DSC$ achieved for the two test datasets is equal to 0.55 and 0.36, respectively.
The authors improve these results in~\cite{meng2019weakly}, developing a CNN-based, weakly supervised method for automatic confidence estimation of shadow regions in 2D US images. By learning and transferring shadow features from weakly-labelled images, continuous shadow confidence maps are directly predicted from input images.
The method is trained and evaluated using a multi-class dataset (8500 US images of 13 fetal standard-planes) with global image-level label (“has shadow” or “shadow-free”) and a single-class dataset (643 fetal brain images). The performances on the 48 multi-class images and on 93 brain images result in $DSC$ equal to 0.54 and 0.71, respectively; the average classification $Acc$ is 0.98. 
Differently from previous work, in which shadow region segmentation is performed, in \cite{meng2019representation} a disentanglement method is presented to disjoint anatomical from shadow features, to generalize anatomical standard plane analysis for abnormality detection in early pregnancy.
The authors proposed a multi-task architecture with adversarial training, evaluated on standard-plane / shadow artifacts classification tasks. The dataset involved contains 8400 US images of 8 fetal standard-planes sampled from 4120 screening examinations.
The $ACC$ achieved for plane recognition is on average 0.94, while for shadow presence identification is about 0.79.
Together with the indiscriminate mixing of image properties, e.g. artifacts and anatomy, another challenge for DL algorithms in US image analysis is the presence of different acquisition devices characteristics.
The work in \cite{meng2020mutual}, while considering the shadow artifacts presence, explores also the cross-device adaptation problem.
In contrast to previous work, the authors present a non-adversarial method that evaluates mutual information between latent features to disentangle categorical features and domain features in a semi-supervised learning framework. 
They evaluate the method on fetal US datasets for two different image classification tasks where domain features are respectively defined by shadow artifacts and image acquisition devices. 
Experiments on the first dataset, consisting of 7000 fetal US images of 6 standard planes sampled from 2694 US scans, aim to separate anatomical and shadow artifacts features (categorical and domain features, respectively) and they result in $Prec$, $Rec$ and $F1$ equal to 0.83, 0.62 and 0.68, respectively. 
Further experiments are performed for a standard plane classification task on a second dataset of 11000 US images acquired from two US devices (different device domains), achieving $Prec$, $Rec$ and $F1$ equal to 0.80, 0.60 and 0.66, respectively.  

\subsubsection{\textbf{Video summarization}}

Video summarization is crucial to lower the workload of clinicians when reviewing US examination.
In \cite{liu2020ultrasound}, DRL is applied to video summarization to select a subset of frames to create a shorter video that contains sufficient and essential information to facilitate retrospective analysis. An encoder-decoder CNN structure is firstly used to extract features from frame sequences; these features are then passed into a bi-directional long short-term memory network (Bi-LSTM) for sequential modeling. After the feature extraction, the summarization task is modeled as a decision-making process using a reinforcement learning (RL) network which, through a reward function, selects frames to be included in the summary. Experiments conducted in a five-fold cross-validation using 10 videos as test set, resulted in a $F1$, $Prec$ and $Rec$ of 0.63, 0.62 and 0.64 respectively.
Video summarization is further investigated in \cite{sharma2021knowledge}, which proposes an approach for video description and  clinical-workflow analysis during fetal US scans. The semi-automatic method to temporally segment the US videos into semantically meaningful segments is based on a double branch architecture in which a SonoNet is employed to extract spatial features, and an LSTM unit is used to model temporal dependency. The segments are then classified into 23 anatomical categories. The resulting semantic annotation is then used to describe operator clinical workflow via machine learning methods, to characterize operator skills and assess operator variability. The automatic annotations of 341 US videos (62 of which manually annotated) resulted in an $Acc$ of 0.92 in four-fold cross-validation.

\subsubsection{\textbf{Probe movement control}}

Accurate obstetric US scanning is highly operator dependent. This issue may be overcome with the use of US probe-movement guidance, especially to assist less-experienced operators. 
In \cite{zhao2021visual}, contrastive learning is used to train an end-to-end self-supervised network for visual-assisted probe movement, via automated landmark retrieval. Firstly, a set of landmarks is built on a virtual 3D fetal model; then, during obstetric scanning, a transformer is used to locate the nearest landmark through descriptor search between the current observation and the landmarks. The global position visualization is displayed on the monitor in real-time as visual guidance for the operator. The model is tested on 10 test cases for a total of 10084 2D US images, achieving an average $Rec$ of 0.94.

\subsubsection{\textbf{Volume reconstruction}}

3D US is extensively used in fetal diagnostic, despite its limited field of view. 3D US volume reconstruction from 2D frames can address the problem, providing more expansive range.
In \cite{luo2021self} a three-fold approach for fetal 3D US reconstruction from 2D slices is proposed. First, the reconstruction process is approached with a differentiable algorithm based on Convolutional LSTM. Second, self-supervised learning is applied on the reconstructed volumes to perform pseudo supervision and regularize the prediction of future frames. Finally, adversarial learning is introduced to improve the representation learning of anatomical shape priors, to prevent uneven reconstructions. The model tested on 13 sequences with 90 2D frames each achieves an $HD$ of 14.12 mm.

\section{Discussion and conclusions}
\label{sec:disc}

\begin{figure}[tbp]
    \centering
    \includegraphics[width = .5\textwidth]{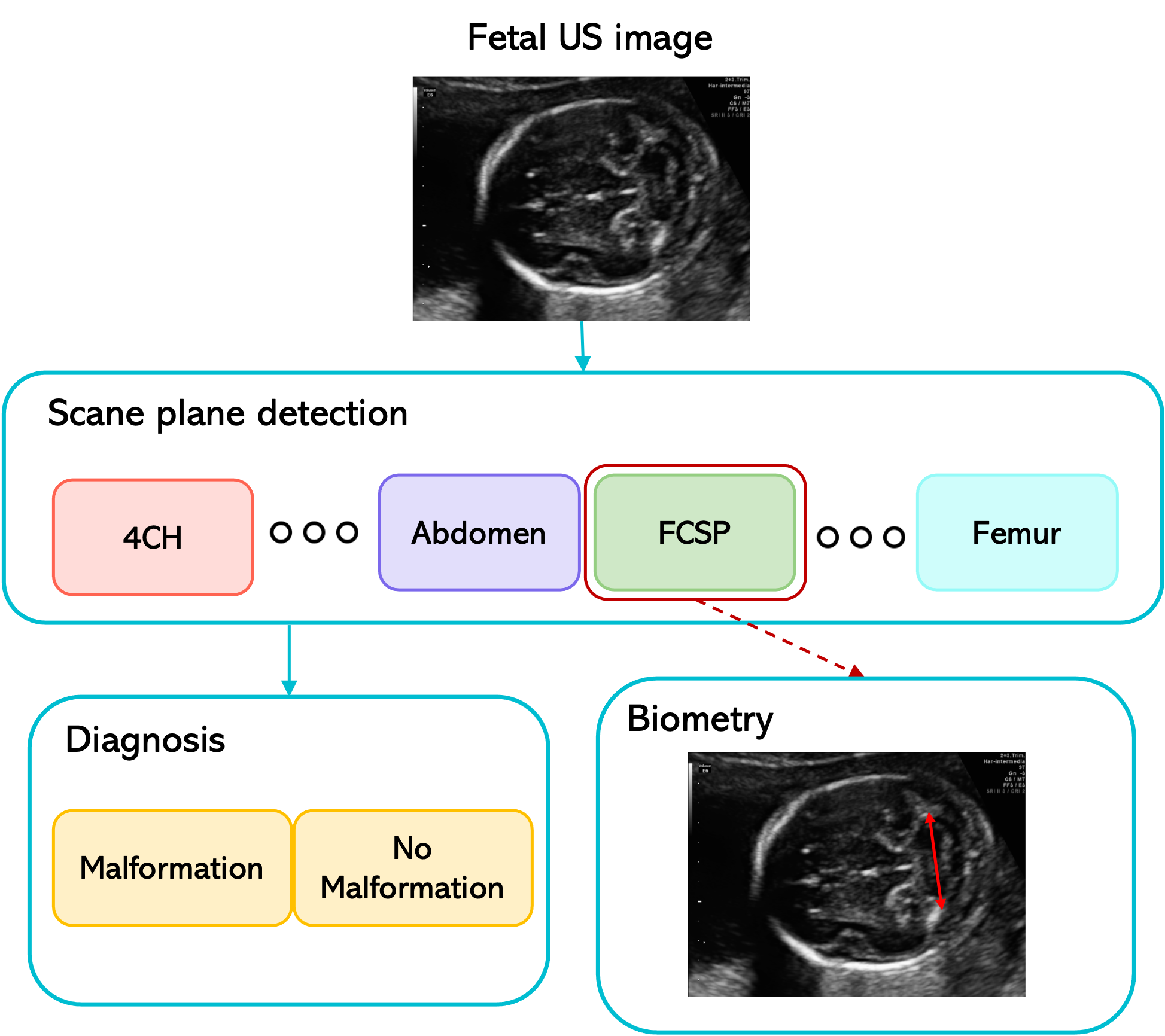}
    \caption{Overall workflow of a computer-assisted tool to support fetal clinicians. }
    \label{fig:ending_figure}
\end{figure}

This review analyzed a wide spectrum of the most recent DL algorithms for fetal US image analysis.
Fetal US image analysis dates back to mid 1900s and a solid and rich literature today exists. 
%
Our survey started with the aim of answering the following questions:

\textbf{Which are the most investigated tasks addressed using DL in the field of fetal US image analysis?}
According to our findings, standard plane detection (19.9\%) and biometry parameter estimation (21.9\%) are among the most investigated tasks (Fig \ref{fig:a}). 
In standard plane detection, fetal brain, abdomen and heart standard planes are the most explored since biometric measurements and identification of abnormalities are mainly performed on these planes. The identification of the specific anatomical landmarks is the key for evaluating the quality of each scanned plane and detection algorithms or attention mechanisms are thus particularly used in this domain. 
As regard biometry parameter estimation, the HC is the most investigated measurement among all the others. This is mainly due to the release of HC18 challenge that allowed a more extensive research in the domain. Segmentation (along with possible head localization pre-processing) followed by fitting methods approaches are the most exploited ones. 
The 46.6\% of the analysed papers dealt with anatomical-structure analysis. According to Fig. \ref{fig:fetal-organ}, the 30,3\% and the 24,2\% of the papers work on fetal heart and brain applications, respectively. The cardiac examination of fetus in literature is designed to maximize the detection of heart anomalies and genetic syndromes. Detection rate is often  optimized by recognizing the main fetal heart planes (4CH, LVOT and RVOT) at first. The analysis is often performed on 2D US images or integrating spatio temporal information, which improves CNN performance since the temporal dependencies present in consecutive frames are exploited. 

As regards fetal brain analysis, a great part of the papers surveyed focuses on the evaluation of brain structures. The analysis is performed both with 2D and, more recently, 3D architectures but structure segmentation is mainly assessed with encoder-decoder inspired architectures. Gestational age and brain development also occupy a consistent part of fetal brain evaluation.
The anatomical regions that have received marginal attention are the lungs, kidney and spine whose approaches are also less varied.

\textbf{Which are the main challenges in regard to fetal examination that are currently tackled by using DL?}
%
From our survey, it emerged that DL is today able to tackle US-image related challenges, including inhomogeneities, artifacts (i.e. shadows), poor contrast and intra- and inter-clinician acquisition and measurement variability, supporting clinicians in a wide range of clinical applications.
For standard plane detection, current DL algorithms are robust when processing single-center datasets, collected and annotated by 1 or 2 experts mostly from a single US machine. The algorithms are able to tackle challenges such as a high variability (size and shape) of fetuses among different GAs, as well as the presence of shared anatomical structures among different standard planes. Here, a winning strategy in the literature is to check that a predefined number of structures appears in the US probe field of view.

As regard anatomical-organ contest, DL has proven to be an essential tool to support the segmentation process of different anatomical structures, which may vary in shape and size over GA. Boundary incompleteness is a common problem for those anatomical structures that are not particularly contrasted with respect to the background. Up to now all the DL segmentation techniques, which are mainly supervised, outperformed other state-of-the art methods in both performance and speed. However, those approaches may be unable to deal with boundary information loss due to a lack of the prior shape information. Dilated convolutions along with residual learning are, thus, often exploited to help increase CNN field of view and recover information loss, respectively. 
Computer-aided diagnosis systems that make full use of DL have been an other research for the past few decades in the field, helping clinicians in their decision-making process. 

For biometry-parameter estimation, DL today achieves robust results on HC estimation tackling challenges such as different position of the head in the image, varying dimensions of fetal head among the gestational trimesters and partially visible head skull. These challenges are today tackled for still images acquired from a single US machine and annotated by a limited number of clinicians. Only one paper proposed an approach in the field invariant to significant image distribution shift between image types \cite{gao2021dual}.
Besides tasks addressed in paragraphs \ref{sec:plane_det},\ref{sec:organs}, \ref{sec:biometry}, recently other tasks are analyzed.
Emerging fields such as preterm birth prediction or fetal pose estimation are investigated to intervene and preserve fetal survival and quantify fetal movements.
Techniques which improve image quality and image analysis are also exploited such as shadow artifact removal and 3D volume reconstruction from 2D slices, which is particularly useful to provide information on the entire anatomy (typical of 3D imaging) together with a more complete field of view.

\textbf{Are the commonly-used datasets sufficient enough for robust DL algorithm development and testing?}
Collecting and sharing annotated datasets for DL algorithm development is a well-known problem for the medical-image analysis community. Up to now, only few datasets are available in the US fetal field (Sec. \ref{sec:dataset}).
Besides practical issues (annotating data is a labour-intensive process), there are social and ethical concerns that have to be considered. To attenuate these issues, all these aspects should be deeply studied in collaboration with non-technical communities (e.g., lawyers, ethicists, ...). This step is crucial with a view to develop robust and generalizable algorithms. 

In terms of number of images, almost all articles were evaluated with less than 1000 images and some of them even with less then 300 images. Moreover, the number of the images that have been used are missing in some of the articles. In order to have a high degree of robustness, train/test size must be always clarified as well as patient number which is an essential information to avoid that the same woman' scan is part of both training/validation and test set.
A cross-validation is strongly suggested when limited data sample is available, especially if we are facing with multi-class problems. With cross-validation strategies all the data is used to evaluate model capability, making algorithm performance more reliable. 

\textbf{Which are open issues that still have to be addressed by DL in the field?}
%
%

Hereafter, we discuss about the main open issues and future research directions that we have identified in the field of fetal US image analysis.


\subsubsection{Multi-expert image annotation}
The lack of image annotation by multiple clinicians currently represents a major drawback.
While it is undeniable that having more than one expert annotator is expensive in term of money and time, this multi-expert annotation is crucial towards the development of robust DL algorithms and the fair comparison of algorithm performance.  
Moreover, the inter-clinician variability could be assessed, which is an indicator of the degree of image complexity. With these considerations in mind, result interpretation must be carefully evaluated when manual annotation is used as the main reference for the study.

\subsubsection{Performance evaluation}
The absence of a systematic evaluation workflow emerged as a critical point. Fair algorithm comparison if often hampered due to an inconsistent use of performance metrics and/ or testing datasets. A direct comparison of methods could not be in fact performed for all the section under exams since each work is validated through different measures. Moreover, despite the efforts of some researchers and International Organizations, the publicly available datasets are still too few, and often released for some specific tasks (e.g., biometric measurements in a body district). 

\subsubsection{Comprehensive analysis}
Developing comprehensive computational models of the fetus is particularly challenging: as the baby grows, the fetus' body is in constant transformation, with significant structural and physiological changes among trimesters leading to an important inter- and intra-organ variability. The creation of anatomically accurate models able to characterize the complexity of fetus anatomy represents, thus, one of the biggest challenges. In addition, the acquisition of quality images is the first step to adequately valuate fetus well-being: an end-to-end approach able to correctly classify fetal planes and successively evaluate biometries or fetus defect hasn't fully exploited yet. Among the few papers that tried to create an unified approach, the work in \cite{arnaout2021ensemble} is the most complete. However, the application is limited to only one anatomical structure (in this case the heart), and thus it does not exploit the potential of DL for distinguishing in-size and shape different objects. 

\subsubsection{Semi, weak and self-supervised learning}
To attenuate the issue of having small annotated datasets, researchers in closer fields are proposing semi, weakly or self-supervised approaches. 
In the fetal US domain, only 13 out 145 papers investigates the potential of such approaches. These include \cite{gao2020label,tan2019semi} to classify and detect planes, and \cite{meng2020mutual} to explore cross-device adaptation problems. Weak supervision tasks are involved in scan plane detection \cite{baumgartner2017sononet}, fetal heart representations \cite{gao2017detection}, multi-organ analysis \cite{gao2019learning} and shadow detection \cite{meng2018automatic,meng2019weakly}. Self supervised techniques are used for scan plane detection \cite{chen2019self}, fetal pose estimation \cite{yang2019fetusmap}, probe movement estimation \cite{zhao2021visual}, fetal 3D reconstruction \cite{luo2021self}.
Other interesting approaches propose unsupervised learning techniques such as in \cite{chen2020cross} for multi-organ classification and in \cite{gao2021dual} for biometry estimation in low cost US devices. 
Even if there is a recent attempt to exploit such techniques in the field, still too few papers makes use of these techniques, proving that this is still an unexplored field in fetal US domain

\subsubsection{Model efficiency}
The computational cost associated to the training and deployment of DL models is not reported in the majority of the surveyed papers. An high computational cost  hampers the efficient deployment of the DL algorithms on single-board computers for on-the-edge computation. At the same time, a high computational cost is associated with a high CO2 consumption. Model efficiency is a critical aspect that is currently monitored by International Organizations \footnote{\url{https://cordis.europa.eu/programme/id/H2020-EU.3.5.}} and performance on this regard should be considered as additional evaluation metric.

\subsubsection{On-device DL in fetal US}
All the world's tech giants embrace DL to provide next-level products and services in fields such social networks, self-driving cars and finance. DL applications in medical imaging give great results in research papers, however, their commercial use is not common yet.
Among the DL algorithms used for commercial purposes in this field, SonoLyst is the first fully integrate AI tool in the world that allows the identification of 20 views recommended by the ISUOG mid-trimester practice guidelines for fetal sonography imaging\footnote{\url{https://www.intelligentultrasound.com/scannav-assist/}} .
 

\subsubsection{Use of federated learning}
The majority of the surveyed papers relies on single-center datasets or datasets made available through international initiatives (e.g., Grand Challenge, Sec.~\ref{sec:dataset}). Accessing sufficiently large and diverse datasets of fetal US images is still a significant challenge. The collaboration among clinical centers based on centrally-shared US images face privacy and ownership concerns. 
Federated learning is a novel paradigm for data-private multi-institutional collaborations, where model-learning leverages all available data without sharing data between institutions~\cite{sheller2020federated}. Despite the benefits of such a paradigm, none of the reviewed articles has implemented it.

\subsubsection{Adherence to the ethics guidelines for trustworthy AI}
With the rapid development of DL algorithms for fetal US image analysis, the application of ethical principles and  guidelines  have  become  crucial. In 2018, the European Commission has published a white paper\footnote{https://ec.europa.eu/futurium/en/ai-alliance-consultation.1.html} titled Ethics Guidelines for Trustworthy AI, to stress on the important of respecting and promoting ethical principles in all the steps that involve the use of DL, from design to deployment. However, no attempts have been made in the reviewed papers in this directions. The Guidelines present an assessment list that offers guidance on each requirement's practical implementation. This assessment list should be filled by researchers and reported as supplementary material. 
\\

To conclude, this review introduced and discussed the most innovative and effective DL methods found in the literature for fetal US image analysis. The methods were summarized using tables reporting performance metrics, training and test set size, and number of annotators.  
Pros and cons of each method were highlighted. 
We hope our review may be helpful for young researchers to get a better picture of the methods available, leading to a speed-up on the development and enhancement of methods for fetal US image analysis.

\newpage

\section*{Nomenclature}

\setlength{\tabcolsep}{6pt} 
\renewcommand{\arraystretch}{1.5} 

 \begin{table}[h!]
    \centering
    \begin{tabularx}{.5\textwidth}{lX}
    
    AC: & Abdominal Circumference
    \\
    $Acc$: & Accuracy \\
    AUC: & Area Under the ROC Curve \\
    AUC-J: & Area Under the ROC Curve by Judd \\
    BPD: & Biparietal Diameter \\
    CNN: & Convolutional Neural Networks \\
    CRL: & Crown-Rump Length \\
    CSP: & Cavum Septum Pellucidum \\
    CTR: & Cardio-Thoracic Ratio \\
    DIFF: & Mean Plane Centres Difference \\
    DL: & Deep Learning \\
    DRL: & Deep Reinforcement Learning \\
    $DSC$: & Dice Similarity Coefficient \\
    $ED$: & Euclidean Distance \\
    FASP: & Fetal Abdomen Standard Plane \\ 
    FBSP: & Fetal Brain Standard Plane \\ 
    FCN: & Fully Convolutional Networks \\
    FCSP: & Fetal Trans-Cerebellum Standard Plane \\
    FFASP: & Fetal Face Axial Standard Plane \\
    FFSP: & Fetal Facial Standard Plane \\
    FFESP: & Fetal Femur Standard Plane \\
    FL: & Femur Diaphysis Length \\
    FLVSP: & Fetal Lumbosacral Spine Standard Plane \\
    FTSP: & Fetal Trans-Thalamic Standard Plane \\
    FVSP: & Fetal Trans-Ventricular Standard Plane \\
    FV: & Fetal Ventriculomegaly \\
    $F1$: & F1-score \\
    GA: & Gestational Age \\
    GAN: & Generative Adversarial Network \\
    GRU-RCN: & Gated-Recurrent-Unit Recurrent Convolutional Network \\
    HC: & Head Circumference \\
    $HD$: & Hausdorff Distance \\
    $IoU$: & Intersection over Union\\
    ISBI: & International Symposium on Biomedical Images \\
    ISUOG: & International Society of Ultrasound in Obstetrics and Gynecology \\
    KLD: & Kullback-Leibler Divergence \\
    LVOT: & Left Ventricular Outflow Tract \\
    LVR: & Lateral Ventricle Ratio \\
    LSTM: & Long Short-Term Memory \\
    $MAE$: & Mean Absolute Error \\
    MICCAI: & Medical Image Computing and Computer Assessed Intervention \\
    MVP: & Maximum Vertical Pocket \\
    NAS: & Neural Architecture Search \\
    NSS: & Normalized Scanpath Saliency \\
    OFD: & Occipito-Frontal Diameter \\ 
    $Prec$: & Precision \\
    $Rec$: & Recall \\
    $RMSE$: & Root Mean Squared Error \\
    \end{tabularx}
    \end{table}
    \begin{table}[h!]
    \centering
    \begin{tabularx}{.5\textwidth}{lX}
    RNN: & Recurrent Neural Network \\
    ROC: & Receiver Operating Characteristic \\
    RVOT: & Right Ventricular Outflow Tract \\
    SE: & Squeeze-and-Excitation \\
    $Spec$: & Specificity \\
    SSL: & Self-Supervised Learning\\ 
    SVM: & Support-Vector Machine \\
    TCD: & Trans Cerebellar Diameter \\
    US: & Ultrasound \\
    3VT: & Three-Vessel Trachea \\
    3VV: & Three-Vessel View \\
    4CH: & Four Chamber View 
    \end{tabularx}
\end{table}


\ifCLASSOPTIONcaptionsoff
  \newpage
\fi

\end{document}